\documentclass[aps,pre,reprint,superscriptaddress,amsmath,amssymb,showpacs,floatfix,longbibliography]{revtex4-1}

\usepackage{graphicx}
\usepackage{dcolumn}
\usepackage{bm}
\usepackage{color}
\usepackage[utf8]{inputenc}
\usepackage{physics}
\usepackage{mathtools}
\usepackage{soul}

\begin{document}
\title{Power-efficiency trade-off for finite-time quantum harmonic Otto heat engine via phase-space approach}
\author{Hyun-Myung Chun}
\affiliation{School of Physics, Korea Institute for Advanced Study, Seoul 02455, Korea}
\author{Jong-Min Park}
\affiliation{Asia Pacific Center for Theoretical Physics, Pohang, 37673, Republic of Korea}
\affiliation{Department of Physics, POSTECH, Pohang 37673, Republic of Korea}

\date{\today}

\begin{abstract}
{
Thermodynamic constraints impose a trade-off between power and efficiency in heat engines, preventing the simultaneous achievement of high power and high efficiency.
For classical microscopic engines, explicit inequalities have been discovered, demonstrating the inherent inevitability of this power-efficiency trade-off.
However, extensions of these results to quantum engines have so far been limited to cases of slow operation.
In this study, we derive a power-efficiency trade-off relation for a paradigmatic quantum engine operating within a finite time, specifically the Otto cycle of a quantum harmonic oscillator. By utilizing a phase-space approach based on quasi-probability representations, we establish a universal trade-off relation applicable to arbitrary time-dependent protocols during the adiabatic processes.
Our results reveal that the power of the quantum engine vanishes as the efficiency approaches the quantum mechanical efficiency bound, which is stricter than the Carnot bound. Furthermore, we identify the conditions under which the upper bound is attained, which indicate maximum power is achieved when the generation of quantum coherence is reduced, and the difference in time durations of the isochoric processes increases.
These findings are validated through numerical calculations, which confirm their applicability across various types of protocols for heat engine cycles.
}
\end{abstract}

\maketitle

\section{Introduction}\label{sec:introduction}

In microscopic systems, traditional thermodynamic concepts often fail to apply.
To explore thermodynamics at microscopic scales, where thermal fluctuations or quantum effects are significant, stochastic thermodynamics~\cite{seifert2012stochastic,peliti2021stochastic} and quantum thermodynamics~\cite{binder2018thermodynamics,strasberg2022quantum} have gained considerable attention. 
Historically, understanding the fundamental principles that enhance heat engine performance has been a central question in thermodynamics and has significantly driven its development.
For instance, the study of the fundamental maximum efficiency of heat engines, known as the Carnot bound, led to the discovery of the second law of thermodynamics.
In this context, investigating universal relations for microscopic heat engines is crucial for advancing our understanding of thermodynamics at microscopic scales.

Heat engine performance is typically characterized by two key factors: power and efficiency.
The maximum efficiency is achieved under reversible and quasi-static operation, which leads to (i) zero power output at the Carnot efficiency and (ii) the efficiency at maximum power being lower than the Carnot efficiency~\cite{curzon1975efficiency,van2005thermodynamic,schmiedl2007efficiency,esposito2009universality,park2016efficiency}.
Despite the long-held belief that this trade-off stemmed from the second law of thermodynamics, recent investigations have shown that microscopic heat engines with broken time-reversal symmetry appear capable of achieving both finite power and the Carnot efficiency simultaneously~\cite{benenti2011thermodynamic}.
This suggests that the second law alone does not fully explain the trade-off between power and efficiency, although these ``dream engines" were ultimately proven impossible due to other stricter conditions~\cite{brandner2013strong,brandner2015thermodynamics,proesmans2015onsager,lee2020exactly}.
In response, various power-efficiency trade-offs have been identified for systems governed by Markov jump processes or Langevin dynamics, providing upper bounds on power that diminish as efficiency approaches the Carnot efficiency~\cite{shiraishi2016universal,dechant2018entropic,pietzonka2018universal}.
These relations demonstrate that even microscopic heat engines are inherently subject to the fundamental trade-off between power and efficiency.

On the other hand, extensive research has explored how quantum effects influence the performance of microscopic heat engines.
For instance, studies have investigated the role of quantum coherence in engine performance, demonstrating that coherence in the reservoirs can enable the engine to surpass the Carnot bound~\cite{scully2003extracting,abah2014efficiency,rossnagel2014nanoscale,niedenzu2016operation},
while internal coherence within the engine itself can be disadvantageous to its performance~\cite{kosloff2002discrete,thomas2014friction,alecce2015quantum,ccakmak2017irreversible,camati2019coherence,lee2020finite}.
These findings suggest that quantum effects could allow systems to exceed the fundamental limits of classical heat engines. To understand the conditions under which this becomes possible, it is crucial to study how quantum effects modify the power-efficiency trade-offs in quantum heat engines.
However, the trade-offs derived so far are limited to engines operating slowly and within the linear response regime~\cite{brandner2016periodic,brandner2017universal,guarnieri2019thermodynamics,brandner2020thermodynamic,miller2021thermodynamic}.

In this Letter, we establish a power-efficiency trade-off for the Otto cycle in quantum harmonic heat engines,
a prototypical model for investigating quantum heat engines~\cite{rezek2006irreversible,rossnagel2014nanoscale,kosloff2017quantum,abah2019shortcut,camati2019coherence,park2019quantum,lee2020finite,singh2020performance,lee2021quantumness,mohanta2023bounds} due to its analytical simplicity and experimental feasibility~\cite{abah2012single,rossnagel2016single,kosloff2017quantum,peterson2019experimental}.
Through a mathematical transformation, we map the quantum harmonic Otto cycle to its classical analogue operating at different effective temperatures.
The mapping allows us to apply methods previously developed for classical heat engines~\cite{dechant2018entropic}, resulting in an upper bound on power.
This bound vanishes as efficiency approaches a quantum mechanical efficiency bound~\cite{park2019quantum},
defined as the Carnot bound in terms of effective temperatures.
Since the quantum mechanical bound is stricter than the Carnot bound for the original temperatures,
our findings reveal that quantum harmonic Otto engines are subject to a tighter trade-off relation than their classical counterparts.
Notably, our trade-off relation applies to any finite-time Otto cycle with arbitrary time-dependent protocols.
We also find that the bound can be attained under specific conditions for the duration of the thermodynamic processes.
Finally, we numerically demonstrate the validity of this bound and attainability across various engine protocols and system parameters.

\section{Model}

We consider a cyclic heat engine consisting of a quantum harmonic oscillator and two heat reservoirs at different temperatures $T_{\rm h}$ and $T_{\rm c}$.
The Hamiltonian of the harmonic oscillator is given by
\begin{equation}
    \hat{H}(t) = \hbar\omega(t) \left( \hat{a}^\dagger(t) \hat{a}(t) + \frac{1}{2} \right),
\end{equation}
where the ladder operators are defined as
\begin{equation}
\begin{aligned}
    \hat{a}(t) & = \sqrt{\frac{m\omega(t)}{2\hbar}} \left( \hat{x} + \frac{i}{m\omega(t)} \hat{p} \right), \\
    \hat{a}^\dagger(t) & = \sqrt{\frac{m\omega(t)}{2\hbar}} \left( \hat{x} - \frac{i}{m\omega(t)} \hat{p} \right),
\end{aligned}
\end{equation}
with frequency $\omega(t)$, mass $m$, position operator $\hat{x}$, and momentum operator $\hat{p}$.

\begin{figure}[t]
\centering
\includegraphics[width=\columnwidth]{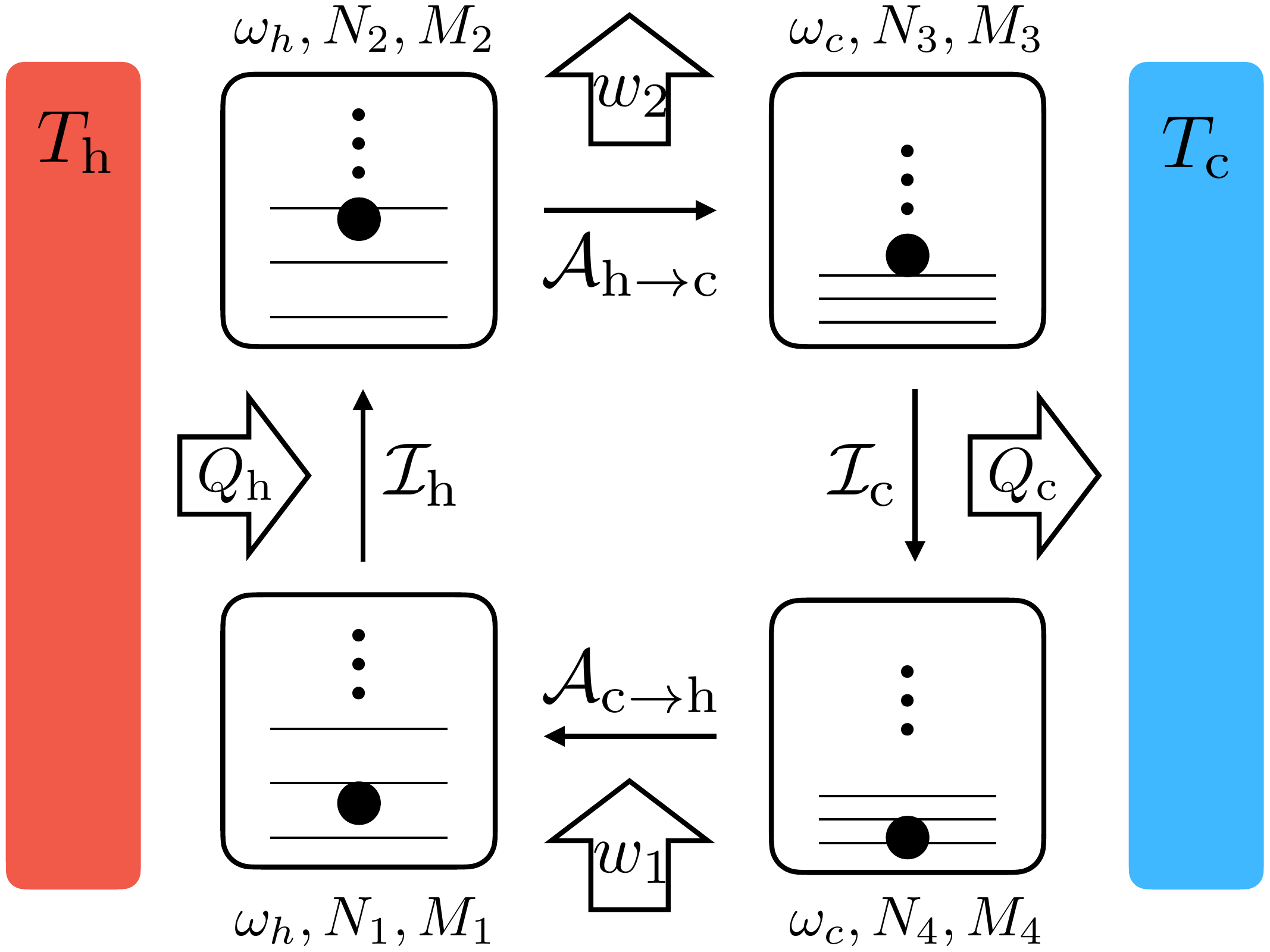}
\caption{A schematic description of a quantum harmonic Otto heat engine.
The horizontal lines in the boxes represent the energy levels of the working substance, and the dots indicate the average energy.
The total extracted work is given by $W_{\rm ext} = w_2 - w_1$.
The quantities $N$ and $M$ are measures of excitation and coherence, respectively, whose definitions are given in the main text.
}
\label{fig:schematic}
\end{figure}

The harmonic oscillator undergoes time evolution according to the Otto cycle, which consists of the following four consecutive processes~\cite{kosloff2017quantum}: (i) isochoric heating ($\mathcal{I}_\textrm{h}$), (ii) adiabatic expansion ($\mathcal{A}_{\textrm{h}\rightarrow \textrm{c}}$), (iii) isochoric cooling ($\mathcal{I}_\textrm{c}$), and (iv) adiabatic compression ($\mathcal{A}_{\textrm{c}\rightarrow \textrm{h}}$), as illustrated in Fig.~\ref{fig:schematic}.

During isochoric processes, the system is in contact with one of the heat reservoirs while the frequency remains constant.
We describe the time evolution of the density operator $\hat{\rho}(t)$ using the Lindblad equation~\cite{breuer2002theory,rezek2006irreversible,camati2019coherence}
\begin{equation}\label{eq:Lindblad_eq}
    \frac{d}{dt}\hat{\rho}(t)
    = -\frac{i}{\hbar}[\hat{H}_{\bullet}, \hat{\rho}(t)]
    + \mathcal{D}_{\bullet}\bm{(}\hat{\rho}(t)\bm{)},
\end{equation}
where $\bullet = \textrm{h}$ or $\textrm{c}$, corresponding to isochoric heating or cooling processes, respectively.
The system Hamiltonian $\hat{H}_\bullet = \hbar \omega_\bullet (\hat{a}^\dagger_\bullet \hat{a}_\bullet + 1/2)$ remains constant, with fixed frequency $\omega_\bullet$ and ladder operator $\hat{a}_\bullet = \sqrt{(m \omega_\bullet)/(2 \hbar)}{\bm (}\hat{x} + i \hat{p}/(m \omega_\bullet){\bm )}$.
The dissipator is a super-operator of the form,
\begin{equation}
\begin{aligned}
    \mathcal{D}_{\bullet}(\hat{\rho})
    & = \gamma_\bullet(\bar{n}_{\bullet} + 1) \left( \hat{a}_{\bullet} \hat{\rho} \hat{a}^\dagger_{\bullet} - \frac{1}{2} \left\{\hat{a}^\dagger_{\bullet} \hat{a}_{\bullet},\hat{\rho} \right\} \right) \\
    & ~~~ + \gamma_\bullet \bar{n}_{\bullet} \left( \hat{a}^\dagger_{\bullet} \hat{\rho} \hat{a}_{\bullet} - \frac{1}{2} \left\{\hat{a}_{\bullet}\hat{a}^\dagger_{\bullet},\hat{\rho} \right\} \right)
\end{aligned}
\end{equation}
with transition rate $\gamma_\bullet$ and the mean excitation of the reservoir $\bar{n}_{\bullet} = (e^{\hbar\omega_\bullet/T_\bullet}-1)^{-1}$.
For brevity, we set $\gamma_\textrm{h} = \gamma_\textrm{c} = \gamma$ and the Boltzmann constant to be unity without loss of generality.

During the adiabatic processes, the system is isolated from the heat reservoirs, with the frequency changing over time between the fixed boundary values $\omega_\textrm{h}$ and $\omega_\textrm{c}$.
The system undergoes unitary evolution, and its dynamics are described by the von Neumann equation
\begin{equation}\label{eq:vN_eq}
    \frac{d}{dt}\hat{\rho}(t)
    = -\frac{i}{\hbar}[\hat{H}(t), \hat{\rho}(t)].
\end{equation}

In Otto engines, work and heat are determined solely by the internal energy $E(t) = \textrm{tr} (\hat{\rho}(t) \hat{H} (t))$ at the boundaries of each processes.
For example, heat absorption $Q_\mathrm{h}$ from the hot reservoir and heat dissipation $Q_\mathrm{c}$ to the cold reservoir during a cycle are given by the internal energy increase during $\mathcal{I}_\mathrm{h}$ and decrease during $\mathcal{I}_\mathrm{c}$, respectively.
Likewise, the net extracted work $W_\mathrm{ext}$ is the sum of internal energy decreases in both adiabatic processes $\mathcal{A}_\mathrm{h \to c}$ and $\mathcal{A}_\mathrm{c \to h}$.
The power and efficiency of the engine are defined as $P = W_\textrm{ext}/\tau$ and $\eta = W_\textrm{ext}/Q_\textrm{h}$, respectively, where $\tau$ is the cycle period.
The Clausius inequality, $-Q_\textrm{h}/T_\textrm{h} + Q_\textrm{c}/T_\textrm{c} \geq 0$, holds for this model, ensuring that the efficiency is bounded by the Carnot bound, $\eta_\mathrm{C} = 1 - T_{\rm c}/T_{\rm h}$~\cite{alicki1979quantum,bender2000quantum,lin2003performance,kieu2004second,kieu2006quantum,gelbwaser2018single}.

\section{Results}

To derive the main result, we use a quasi-probability representation, where the system state is described by a real-valued function of phase variables, known as the Wigner function~\cite{puri2001mathematical,gardiner2004quantum}.
Using the eigenvalue $\alpha$ of the annihilation operator $\hat{a}$ as the phase variable, we map the time-evolution equations \eqref{eq:Lindblad_eq} and \eqref{eq:vN_eq} to a Fokker-Planck equation for the complex-valued variables $\alpha$ and $\alpha^*$ (see Sec.~\ref{sec:derivation} for details).
A proper variable transformation further enables us to interpret the dynamics of isochoric processes as those of a classical harmonic oscillator driven by a nonconservative force but at different temperatures $\tilde{T}_\bullet = \hbar \omega_\bullet ( \bar{n}_\bullet + 1/2 )$ (See the Supplemental Material~\cite{SM}).
\nocite{dechant2018entropic,oono1998steady,esposito2010three,maes2014nonequilibrium,dechant2022geometric,hatano2001steady,oono1998steady,gardiner2009stochastic,puri2001mathematical,gardiner2004quantum,cahill1969ordered,cahill1969density,puri2001mathematical,puri2001mathematical,gardiner2004quantum,kosloff2002discrete,thomas2014friction,alecce2015quantum,ccakmak2017irreversible,rezek2006irreversible,kosloff2017quantum,lee2020finite,lee2021quantumness}
The stochastic thermodynamic definition of heat in this effective classical system reproduces the same value as the original quantum mechanical definition~\cite{santos2017wigner,park2019quantum}.
The connection between our approach and the entropic bound derived in \cite{dechant2018entropic}, the latter being valid for classical systems described by stochastic thermodynamics, is discussed further in the Supplemental Material~\cite{SM}.

Using this mapping and applying methods inspired by\cite{dechant2018entropic}, as presented in Sec.~\ref{sec:derivation}, we derive the following power-efficiency trade-off
\begin{equation}\label{eq:trade-off}
    P \leq 
    \gamma \tilde{T}_\textrm{h} \tilde{\phi}_\textrm{min}^\mathcal{I}
    \frac{ \eta(\tilde{\eta} - \eta)}{1 - \eta},
\end{equation}
where $\tilde{\phi}^\mathcal{I}_\textrm{min}$ represents the fraction of the shorter isochoric process duration to the total cycle period, and $\tilde{\eta}=1-\tilde{T}_\textrm{c}/\tilde{T}_\textrm{h}$ is a quantum mechanical efficiency bound~\cite{park2019quantum}, satisfying $\eta \leq \tilde{\eta} \leq \eta_\textrm{C}$.
This relation exhibits several distinct differences from existing power-efficiency trade-offs.
First, the bound vanishes as the efficiency approaches the quantum mechanical bound $\tilde{\eta}$, rather than the Carnot bound.
Second, the prefactor is determined by a few quantities with clear physical interpretations.
For instance, $\tilde{T}_\mathrm{h}$ represents the internal energy of the harmonic oscillator in equilibrium with the hot reservoir.
The prefactor also implies that the engine cannot generate power if either istochoric process is absent, i.e., $\tilde{\phi}^\mathcal{I}_\textrm{min}=0$.
Lastly, the $\eta$-dependence of the bound is different from that of existing bounds, which typically take the form $\chi \eta (\eta_\mathrm{C} - \eta)$ with a system parameter-dependent prefactor $\chi$~\cite{shiraishi2016universal,dechant2018entropic,pietzonka2018universal,kwon2024unified}.
A similar relation can be obtained for the classical version of engines as shown in the Supplemental Material~\cite{SM}.
The maximum bound with respect to $\eta$ is attained at $\eta = \tilde{\eta}_{\rm CA} \equiv 1 - \sqrt{\tilde{T}_\textrm{c}/\tilde{T}_\textrm{h}}$, resulting in $P \leq \gamma \tilde{T}_{\rm h} \tilde{\phi}_{\rm min}^\mathcal{I} \tilde{\eta}_{\rm CA}^2$.
In the high-temperature limit, $\tilde{\eta}_{\rm CA}$ converges to the Curzon-Ahlborn efficiency~\cite{curzon1975efficiency}, which is consistent with the previous results~\cite{rezek2006irreversible,deffner2018efficiency}.

We find that the power bound is saturated under two conditions.
First, no coherence is generated during the cycle.
Second, the duration of one of the isochoric processes is much longer than that of the other.
One possible setup satisfying the first condition is a cycle with quasi-static adiabatic processes, as justified by the quantum adiabatic theorem~\cite{kato1950adiabatic}.
For the Otto cycle with quasi-static adiabatic processes, the power is given by~\cite{rezek2006irreversible, park2019quantum}
\begin{equation}\label{eq:power_quasistatic}
    P_q = \tilde{T}_\mathrm{h}\frac{\eta(\tilde{\eta}-\eta)}{\tau(1-\eta)}
    \frac{(1 - e^{-\gamma \Delta t(\mathcal{I}_{\rm h})})(1 - e^{-\gamma \Delta t(\mathcal{I}_{\rm c})})}{1 - e^{-\gamma \bm{(} \Delta t(\mathcal{I}_{\rm h}) + \Delta t(\mathcal{I}_{\rm c}) \bm{)} }}
\end{equation}
where $\Delta t(\mathcal{I}_\bullet)$ is the duration of the isochoric process $\mathcal{I}_\bullet$.
The bound in \eqref{eq:trade-off} is saturated when $ \min\{ \Delta t(\mathcal{I}_{\rm h}), \Delta t(\mathcal{I}_{\rm c}) \} \ll \gamma^{-1} \ll \max \{ \Delta t(\mathcal{I}_{\rm h}), \Delta t(\mathcal{I}_{\rm c}) \}$, which satisfies the second condition of saturation.
However, when the maximum power is attained, the actual amount of extracted work per cycle is negligibly small because $\gamma \tilde{\phi}^\mathcal{I}_\textrm{min} \ll 1/\tau$.

Since we provide no proof that these conditions are also necessary, other conditions may exist under which the bound is attained with finite extracted work.
Nevertheless, in the following numerical analysis, we show that for various protocols the bound is saturated only when the conditions we identified are satisfied. Details of the equality conditions and the derivation of the power of quasi-static engines are presented in the Supplemental Material~\cite{SM}.

\begin{figure}[t]
\centering
\includegraphics[width=\columnwidth]{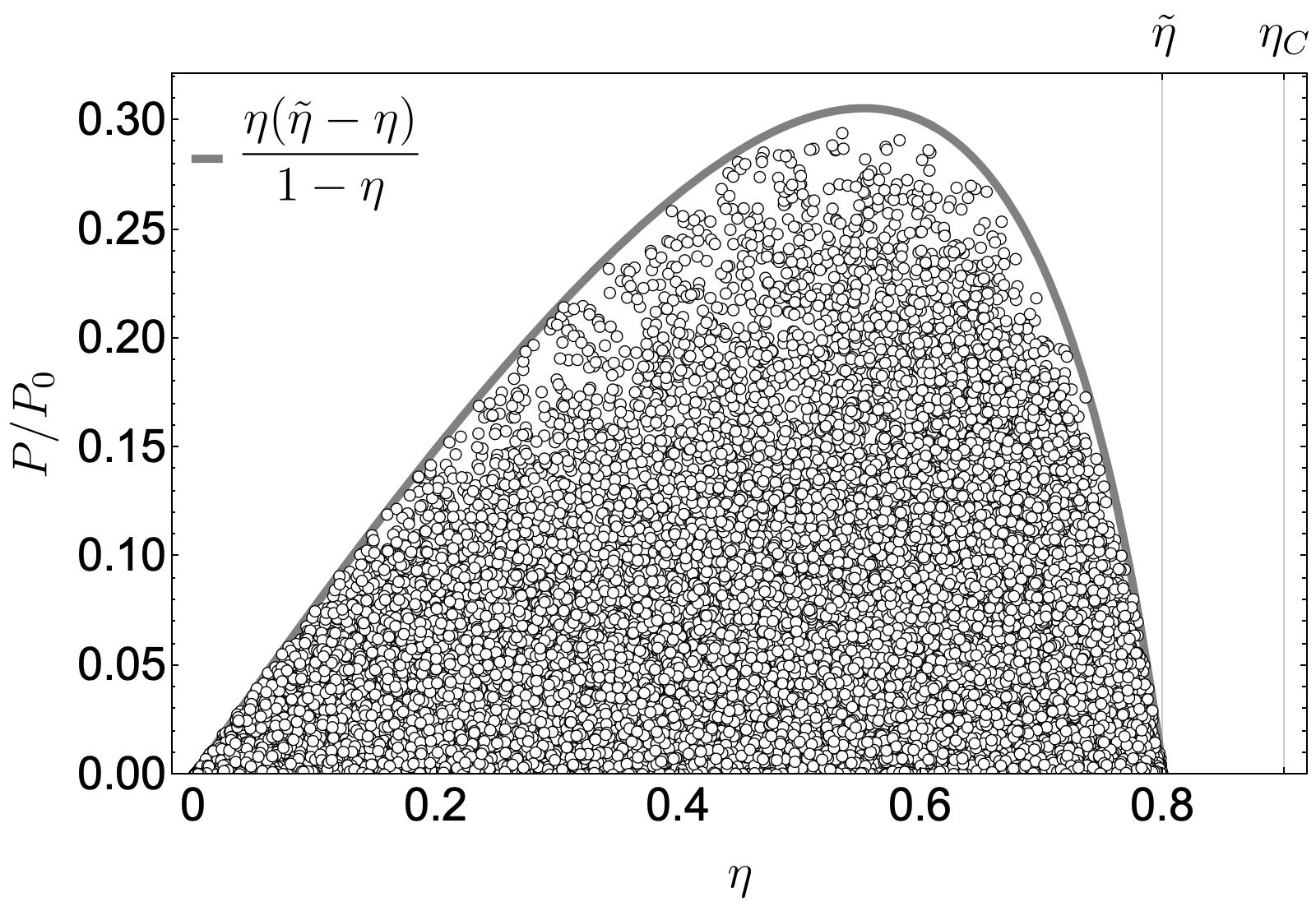}
\caption{The validity of the power-efficiency trade-off in \eqref{eq:trade-off} is demonstrated, where $P_0 \equiv \gamma \tilde{T}_{\rm h} \tilde{\phi}^\mathcal{I}_{\rm min}$ denotes the prefactor on the right-hand side.
Each circle represents the power and efficiency of a harmonic Otto heat engine with a randomly selected set of system parameters.
The number of circles is $10^5$.}
\label{fig:illustration}
\end{figure}

\begin{figure}[t]
\centering
\includegraphics[width=\columnwidth]{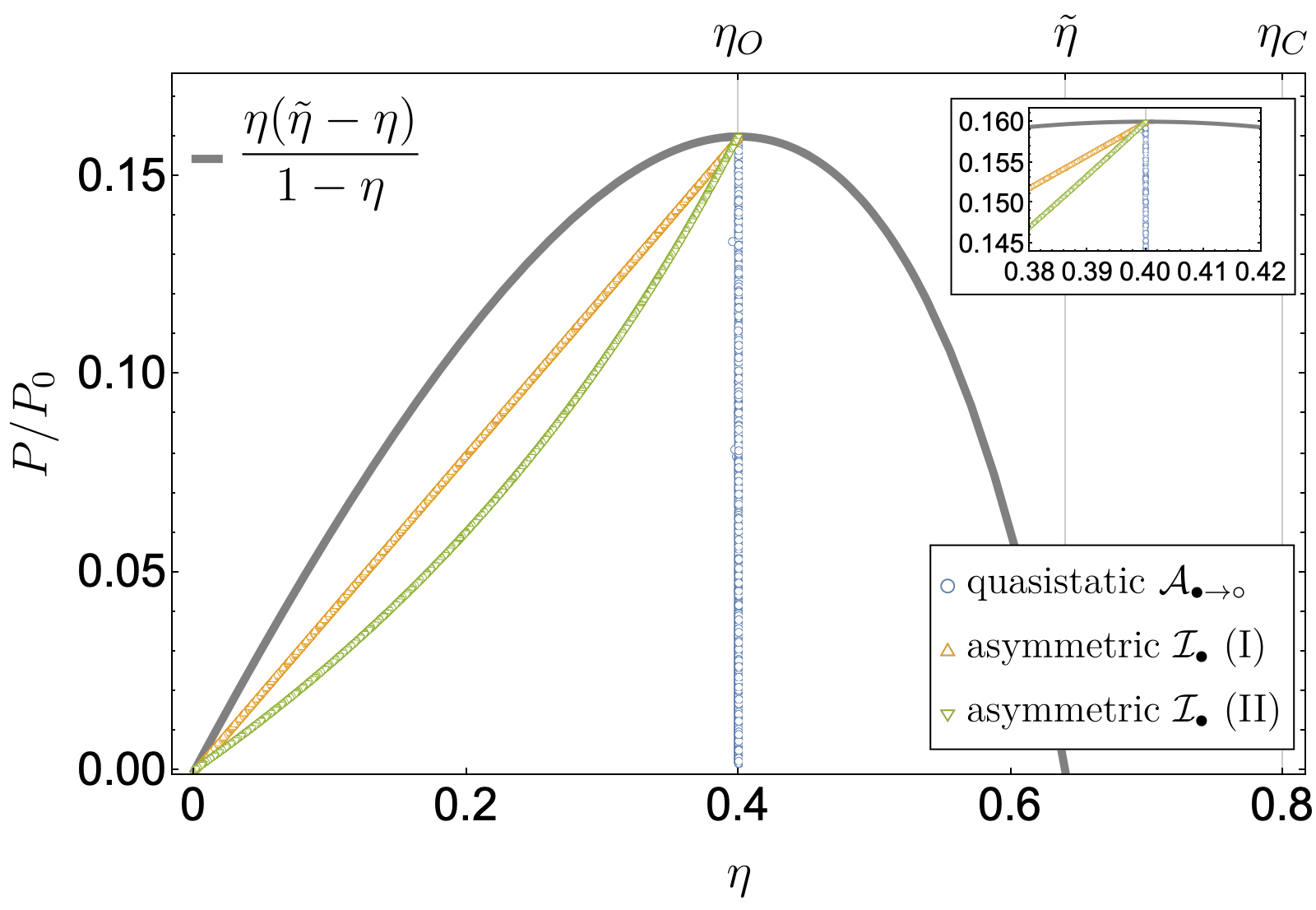}
\caption{The saturation conditions of the power-efficiency trade-off in \eqref{eq:trade-off} are examined, where $P_0$ denotes the prefactor on the right-hand side.
Each symbol represents the power and efficiency of a harmonic Otto heat engine at extreme values for process durations.
Circles correspond to engine cycles with quasi-static adiabatic processes, while triangles correspond to engine cycles with asymmetric isochoric processes.
The Otto efficiency, the quantum mechanical bound, and the Carnot efficiency are fixed.
Other independent system parameters are randomly sampled.
The number of each type of symbols is $10^4$.}
\label{fig:saturation}
\end{figure}

\section{Simulation}

We demonstrate our findings through numerical simulations of quantum engines with various protocols. At the periodic steady state, the system remains in a Gaussian state and is fully characterized by the following second moments: $N(t) = \langle (a^\dagger a + a a^\dagger)/2 \rangle$, $M(t) = \langle a^2 \rangle$, and $M^*(t) = \langle (a^\dagger)^2 \rangle$, where $\langle \hat{o} \rangle = {\rm tr}\{\hat{\rho} \hat{o}\}$ with any operator $\hat{o}$. Furthermore, their time evolution is governed by a closed set of linear equations, enabling the internal energy and the energy flows to be determined by numerically integrating these coupled equations.

We first focus on the protocol satisfying $\dot{\omega}(t) \propto \omega^2(t)$ during both adiabatic processes, as this protocol is analytically solvable and often used in the literature~\cite{kosloff2017quantum,lee2020finite,lee2021quantumness}. Figure~\ref{fig:illustration} illustrates the validity of the power-efficiency trade-off \eqref{eq:trade-off} for this specific protocol.
The symbols represent numerical data obtained for randomly selected system parameters.
We randomly sampled the durations $\Delta t(\mathcal{P})$ of processes for $\mathcal{P} \in \{ \mathcal{I}_{\rm h}, \mathcal{I}_{\rm c}, \mathcal{A}_{{\rm h} \rightarrow {\rm c}},  \mathcal{A}_{{\rm c} \rightarrow {\rm h}}\}$, $\gamma$, $\hbar \omega_\mathrm{h}$, and $\beta_\mathrm{h} = 1/T_\mathrm{h}$ from a uniform distribution over the range $[0,10]$, under the conditions $\eta_\mathrm{C} = 0.9$ and $\tilde{\eta} = 0.8$. 
All the data points lie within the bound in \eqref{eq:trade-off}, with some points close to the bound, implying that the bound can be saturated for the protocol satisfying $\dot{\omega}(t) \propto \omega^2(t)$.
Additionally, we perform numerical simulations for more general protocols satisfying $\dot{\omega}(t) \propto \omega^m(t)$ with $m = -2, -1, 0, 1,$ and $2$.
As shown in the Supplemental Material~\cite{SM}, the results consistently align with our predictions for all combinations of $m$ for each isochoric process.

To examine the saturation conditions for the power-efficiency trade-off, we calculate the power and efficiency of the heat engine with extreme values for durations $\Delta t(\mathcal{P})$.
For the first condition for saturation, we consider engine cycles with slow adiabatic processes by setting $\Delta t(\mathcal{A}_{{\rm h}\to {\rm c}}) = \Delta t(\mathcal{A}_{{\rm c}\to {\rm h}}) = 1000 \max \{ \Delta t(\mathcal{I}_{\rm h}), \Delta t(\mathcal{I}_{\rm c}) \}$, where the durations of the isochoric processes are randomly sampled from the range $[0,10]$.
The circle symbols in Fig.~\ref{fig:saturation} represent the resulting power and efficiency for randomly selected system parameters, with efficiencies tightly concentrated near the quasistatic efficiency, often referred to as the Otto efficiency $\eta_O = 1 - \omega_{\rm c}/\omega_{\rm h}$~\cite{rezek2006irreversible,camati2019coherence}.
For the second condition for saturation, we consider engine cycles with asymmetric isochoric processes by setting $\min\{ \Delta t(\mathcal{I}_{\rm h}), \Delta t(\mathcal{I}_{\rm c}) \} = 0.001 \gamma^{-1}$ and $\max\{ \Delta t(\mathcal{I}_{\rm h}), \Delta t(\mathcal{I}_{\rm c}) \} = 1000 \gamma^{-1}$.
The durations of adiabatic processes are randomly sampled from the range $[0,10]$.
The triangle symbols in Fig.~\ref{fig:saturation} represent the resulting power and efficiency, which are narrowly distributed along two curves: the upright triangles correspond to engine cycles with $\Delta t(\mathcal{I}_{\rm h}) \ll \Delta t(\mathcal{I}_{\rm c})$, while the reversed triangles correspond to the opposite scenario.
In both cases, $\hbar\omega_{\rm h}$ and $\gamma$ are randomly sampled from the range  $[0,10]$.
The parameters $\hbar\omega_c$, $\beta_h$, and $\beta_c$ are determined from the conditions that $\eta_{\rm C} = 0.8$, $\tilde{\eta} = 0.64$, and $\eta_O = 0.4$.
Figure~\ref{fig:saturation} illustrates that the trade-off is saturable when both conditions are satisfied.
The numerical validation for other protocols are presented in the Supplemental Material~\cite{SM}.

\section{Derivation}\label{sec:derivation}

In this section, we derive the main result by using the quasi-probability representation, which has proven useful for leveraging the tools of stochastic thermodynamics by mapping quantum dynamics to classical stochastic dynamics~\cite{santos2017wigner,park2019quantum}.
In this approach, the coherent basis is adopted to define the quasi-probability distributions.
A coherent state $|\alpha\rangle$ is an eigenstate of the annihilation operator with a complex eigenvalue $\alpha$, i.e., $\hat{a}|\alpha\rangle = \alpha|\alpha \rangle$.
The set of coherent states forms a non-orthogonal overcomplete basis~\cite{gardiner2004quantum}.

Quasi-probability distributions are defined through the inverse Fourier transform of a moment-generating function in analogy to classical cases. One typical example is the Wigner function~\cite{puri2001mathematical,gardiner2004quantum}, which is defined by
\begin{equation}\label{eq:def_Wigner_dist}
    f(\alpha,\alpha^*,t)
    = \frac{1}{\pi^2} \int {\rm tr} \left( \hat{\rho}(t) e^{-i \bm{(}\xi\hat{a}(t)+\xi^*\hat{a}^\dagger(t)\bm{)} } \right) e^{i(\xi\alpha + \xi^*\alpha^*)} d^2\xi .
\end{equation}
where the integral $\int d^2\xi$ is taken over the complex plane.
Unlike classical cases, the non-commutativity of quantum operators necessitates the selection of a specific ordering of exponentiated operators, yielding indefinitely many kinds of quasi-probability distributions corresponding to any specific ordering.
The Wigner function in~\eqref{eq:def_Wigner_dist} corresponds to the symmetric ordering. In this Letter, we focus on this specific representation and the discussion of general cases is addressed in the Supplemental Material~\cite{SM}.

With the quasi-probability representation, one can describe the dynamics of the system during each process of the Otto cycle as the time evolution of the Wigner function corresponding to that of the density operator.
Using the definition in \eqref{eq:def_Wigner_dist} and integration by parts, which is justified under the condition that $f(\alpha,\alpha^*,t)$ vanishes rapidly at $|\alpha|\to \infty$, one can show that the time evolution of the Wigner function is expressed by the following Fokker-Planck equation for complex-valued variables during isochoric processes~\cite{santos2017wigner,gardiner2004quantum}:
\begin{equation}\label{eq:complex_FP_eq_isochoric}
\begin{aligned}
    \frac{\partial}{\partial t} f(\alpha,\alpha^*,t)
    & = \mathcal{L}_\bullet f(\alpha,\alpha^*,t) \\
    & ~~~ + \frac{\partial J_{\bullet}(\alpha,\alpha^*,t)}{\partial \alpha}
    + \frac{\partial J^*_{\bullet}(\alpha,\alpha^*,t)}{\partial \alpha^*},
\end{aligned}
\end{equation}
where
\begin{equation}
    \mathcal{L}_\bullet
    = -i\omega_\bullet \left(-\frac{\partial}{\partial \alpha} \alpha
    + \frac{\partial}{\partial \alpha^*} \alpha^* \right)
\end{equation}
is a differential operator and
\begin{equation}\label{eq:phase_current}
\begin{aligned}
    J_{\bullet}(\alpha,\alpha^*,t)
    & = \frac{\gamma \alpha}{2} f(\alpha,\alpha^*,t) \\
    & ~~~ + \frac{\gamma}{2} \left( \bar{n}_\bullet + \frac{1}{2} \right)
    \frac{\partial f(\alpha,\alpha^*,t)}{\partial \alpha^*}
\end{aligned}
\end{equation}
is a complex-valued phase-space current.
Similarly, the time evolution during adiabatic processes is given by
\begin{equation}\label{eq:complex_FP_eq_adiabatic}
    \frac{\partial}{\partial t} f(\alpha,\alpha^*,t)
    =  \mathcal{L}_{\mathcal{A}}(t) f(\alpha,\alpha^*,t)
\end{equation}
where
\begin{equation}\label{eq:L_A}
\begin{aligned}
    \mathcal{L}_{\mathcal{A}}(t)
    & = -i\omega(t) \left(-\frac{\partial}{\partial \alpha} \alpha
    + \frac{\partial}{\partial \alpha^*} \alpha^* \right) \\
    & ~~~ - \frac{\dot{\omega}(t)}{2\omega(t)} \left( 
    \alpha^* \frac{\partial}{\partial \alpha} 
    + \alpha \frac{\partial}{\partial \alpha^*} \right).
\end{aligned}
\end{equation}

In addition, one can express the thermodynamic quantities in terms of averaged quantities over the Wigner function.
Using the identity $e^{-i(\xi\hat{a} + \xi^* \hat{a}^\dagger)} = e^{-|\xi|^2/2} e^{-i\xi^*\hat{a}^\dagger} e^{-i\xi\hat{a}}$, the internal energy can be identified as
\begin{equation}
    E(t) = {\rm tr}{\bm (} \hat{\rho}(t)\hat{H}(t){\bm )} = \hbar{\omega}(t)\int |\alpha|^2 f(\alpha,\alpha^*,t) d^2\alpha.
\end{equation}
As mentioned in the main text, the change in energy is solely attributed to either heat flow (during isochoric processes) or work extraction (during adiabatic processes), allowing us to determine the rates of heat and work by taking the time derivative of $E(t)$.
Consequently, the rates of heat and work are given by
\begin{equation}\label{eq:heat_Wigner}
\begin{aligned}
    \dot{Q}_\bullet(t)
    & = - \hbar \omega_\bullet \int \bm{(} \alpha^* J_\bullet(\alpha,\alpha^*,t)
    + \alpha J_\bullet^*(\alpha,\alpha^*,t) \bm{)} d^2\alpha \\
    & = \gamma \hbar \omega_\bullet \left(\bar{n}_\bullet + \frac{1}{2} - \int |\alpha|^2 f(\alpha,\alpha^*,t) d^2 \alpha \right),
\end{aligned}
\end{equation}
\begin{equation}\label{eq:work_Wigner}
    \dot{W}(t) = \frac{\hbar \dot{\omega}(t)}{2} \int (\alpha + \alpha^*)^2 f(\alpha,\alpha^*,t) d^2\alpha,
\end{equation}
respectively.

Despite the possible issue that the Wigner function can take on negative values~\cite{puri2001mathematical,gardiner2004quantum}, the Wigner function of harmonic Otto heat engines remains non-negative, allowing us to define its Shannon entropy as
\begin{equation}
    S(t) = -\int f(\alpha,\alpha^*,t) \ln f(\alpha,\alpha^*,t)  d^2\alpha.
\end{equation}
This is because the time-evolution equations in \eqref{eq:complex_FP_eq_isochoric} and \eqref{eq:complex_FP_eq_adiabatic} correspond to the Ornstein-Uhlenbeck process, ensuring that the Wigner function retains a Gaussian form throughout the engine cycle at the cyclic steady state.
From the perspective of stochastic thermodynamics~\cite{seifert2012stochastic}, we may regard the Shannon entropy of the Wigner function as an effective system entropy and define an effective total entropy production rate as~\cite{santos2017wigner}
\begin{equation}\label{eq:tot_EP}
\begin{aligned}
    \dot{\Pi}_\bullet(t) 
    & = \dot{S}(t) - \frac{\dot{Q}_\bullet(t)}{\tilde{T}_\bullet} \\
    & = \frac{4}{\gamma(\bar{n}_\bullet+\frac{1}{2})}  \int \frac{|J_{\bullet}(\alpha,\alpha^*,t)|^2}{f(\alpha,\alpha^*,t)}  d^2 \alpha \geq 0
\end{aligned}
\end{equation}
for isochoric processes, with the effective temperature  $\tilde{T}_{\bullet} = \hbar\omega_\bullet ( \bar{n}_\bullet + 1/2 ) = (\hbar\omega_\bullet/2) \coth{\bm (} \hbar\omega_\bullet/(2 T_\bullet){\bm )}$.
We remark that the effective temperature converges to the original temperature $T_\bullet$ when the quantum energy gap is much smaller than the thermal energy, i.e., $ \hbar \omega_\bullet \ll T_\bullet$.
One can also see that $\dot{S}(t) = 0$ during adiabatic processes, resulting in the Clausius inequality
$-Q_\textrm{h}/\tilde{T}_\textrm{h} + Q_\textrm{c}/\tilde{T}_\textrm{c} \geq 0$. Thus, it follows that the efficiency is constrained by a quantum mechanical upper bound, $\eta = W_{\rm ext}/Q_\textrm{h} \leq \tilde{\eta} \equiv 1 - \tilde{T}_\textrm{c}/ \tilde{T}_\textrm{h}$~\cite{park2019quantum}.

We apply the Cauchy-Schwarz inequality to \eqref{eq:heat_Wigner}, noting that heat absorption $Q_\textrm{h}$ has the form of an inner product $(\bm{A},\bm{B}) = \int_{\mathcal{I}_\textrm{h}}  dt \int d^2\alpha \bm{A}(\alpha,\alpha^*,t) \cdot \bm{B}^*(\alpha,\alpha^*,t)$
of complex column vectors $\bm{A} = -\hbar\omega_\textrm{h} \sqrt{f} (\alpha, \alpha^*)^{\rm T}$ and $\bm{B} = (1/\sqrt{f}) (J_\textrm{h},J_\textrm{h}^*)^{\rm T}$, where the time integral is taken over the isochoric process $\mathcal{I}_\textrm{h}$.
The resulting inequality is
\begin{equation}\label{eq:Cauchy-Schwartz}
\begin{aligned}
    Q_\textrm{h}^2 
    & \leq \left( 2\hbar^2 \omega_\textrm{h}^2 \int_{\mathcal{I}_\textrm{h}} dt \int d^2\alpha ~ |\alpha|^2 f(\alpha,\alpha^*,t)   \right) \\
    & ~~~ \times \left(
    2\int_{\mathcal{I}_\textrm{h}} dt \int d^2 \alpha ~ \frac{|J_\textrm{h}(\alpha,\alpha^*,t)|^2}{f(\alpha,\alpha^*,t)} \right) \\
    & = \tilde{T}_\textrm{h} \Pi_\textrm{h}  
    \left( \gamma \tilde{T}_\textrm{h} \Delta t(\mathcal{I}_\textrm{h}) - Q_\textrm{h} \right),
\end{aligned}
\end{equation}
where \eqref{eq:heat_Wigner} and \eqref{eq:tot_EP} have been used in the equality, and $\Pi_\textrm{h} = \int_{\mathcal{I}_\textrm{h}} \dot{\Pi}_\textrm{h} (t) dt$.
Using the relation $\Pi_\textrm{h} \leq \Pi_\textrm{h} + \Pi_\textrm{c} = -Q_\textrm{h}/\tilde{T}_\textrm{h} + Q_\textrm{c}/\tilde{T}_\textrm{c} = (1/\tilde{T}_\textrm{c} - 1/\tilde{T}_\textrm{h}) Q_\textrm{h} -  W_{\rm ext}/\tilde{T}_\textrm{c}$ and the definitions of power $P = W_{\rm ext}/\tau$ and efficiency $\eta = W_{\rm ext}/Q_\textrm{h}$, we derive a power-efficiency trade-off,
\begin{equation}\label{eq:trade-off_h}
    P \leq \frac{\gamma \tilde{T}_\textrm{h} \Delta t(\mathcal{I}_\textrm{h})}{\tau}
    \frac{ \eta(\tilde{\eta} - \eta)}{1 - \eta}
\end{equation}
with $\tilde{\eta} = 1 - \tilde{T}_\textrm{c}/\tilde{T}_\textrm{h}$.
Similarly, applying the Cauchy-Schwartz inequality to $Q_\textrm{c}$ and using the inequality $\Pi_\textrm{c} \leq \Pi_\textrm{h} + \Pi_\textrm{c}$, we can obtain another trade-off with the same form as \eqref{eq:trade-off_h} but with $\Delta t(\mathcal{I}_\textrm{c})$ replacing $\Delta t(\mathcal{I}_\textrm{h})$.
Combining the two bounds on power, we have a tighter power-efficiency trade-off,
\begin{equation}\label{eq:trade-off2}
    P \leq \gamma \tilde{T}_\textrm{h} \frac{\min \{ \Delta t(\mathcal{I}_\textrm{h}), \Delta t(\mathcal{I}_\textrm{c}) \} }{\tau}
    \frac{ \eta(\tilde{\eta} - \eta)}{1 - \eta}.
\end{equation}
Denoting $\tilde{\phi}_{\rm min}^{\mathcal{I}} = \min \{ \Delta t(\mathcal{I}_\textrm{h}), \Delta t(\mathcal{I}_\textrm{c}) \} / {\tau}$, we arrive at the main result in \eqref{eq:trade-off}.

\section{Conclusion}

We investigated the trade-off relation between power and efficiency in the Otto cycle of a quantum harmonic oscillator. Using a quasi-probability representation, we derived an upper bound on the power for finite-time quantum Otto engines, which can be regarded as a variant of the power-efficiency trade-off relation. The relation reveals that the power of quantum heat engines is constrained by a stricter efficiency bound compared to classical heat engines. This suggests that quantum Otto engines are unlikely to achieve higher performance than their classical counterparts.

Furthermore, we identified the conditions under which the bound is saturated. One condition highlights the detrimental impact of coherence on the power of quantum Otto engines. Another condition indicates that a significant difference in time durations between the two isochoric processes may lead to enhanced power.  These findings deepen our understanding of the fundamental principles governing the performance of quantum heat engines.

This work demonstrates the potential for extending the universal relations for classical microscopic heat engines to quantum heat engines using quasi-probability representation. It is promising that a similar approach is applicable to two-level systems~\cite{santos2018spin,zicari2023role} or thermodynamic processes involving nonequilibrium thermal reservoirs such as  squeezed thermal reservoirs or dephasing reservoirs~\cite{rossnagel2014nanoscale,santos2017wigner,klaers2017squeezed}. Extending this approach to other quantum heat engines would be valuable future work.

\begin{acknowledgments}
\emph{Acknowledgments} -- H.-M.C. was supported by a KIAS Individual Grant (PG089401) at Korea Institute for Advanced Study.
J.-M.P. was supported by an appointment to the JRG Program at the APCTP through the Science and Technology Promotion Fund and Lottery Fund of the Korean Government, also by the Korean Local Governments - Gyeongsangbuk-do Province and Pohang City.
\end{acknowledgments}

\appendix
\bibliographystyle{apsrev}
\bibliography{main}

\end{document}


\title{Supplemental Material --- Power-efficiency trade-off for finite-time quantum harmonic Otto heat engine via phase-space approach}
\author{Hyun-Myung Chun}
\affiliation{School of Physics, Korea Institute for Advanced Study, Seoul 02455, Korea}
\author{Jong-Min Park}
\affiliation{Asia Pacific Center for Theoretical Physics, Pohang, 37673, Republic of Korea}
\affiliation{Department of Physics, POSTECH, Pohang 37673, Republic of Korea}

\maketitle

\section{Mapping to effective classical systems}\label{sec:app1}
In this section, we discuss an effective classical system equivalent to the one described by the Fokker-Planck equation for the complex-valued variables $\alpha$ and $\alpha^*$.
This discussion highlights structural similarities between our results and existing relations for classical systems within the framework of stochastic thermodynamics.
To this end, we introduce the variable transformation $\alpha = (x + i y)/\sqrt{2}$, where $x$ and $y$ are real-valued. This transformation ensures that the Jacobian determinant remains unity, so that $f(\alpha, \alpha^*, t) = f(x, y, t)$. The Fokker-Planck equation for the isochoric process, in Eqs.~(9-11) in the main text, is then mapped to
\begin{equation}\label{eqA:real_FP_eq}
    \dot{f} (\boldsymbol{x},t)
    = \boldsymbol{\nabla}^{\rm T}
    \cdot \left( 
    \begin{pmatrix}
        \frac{\gamma}{2} & - \omega \\
         \omega &\frac{\gamma}{2}
    \end{pmatrix}
    \boldsymbol{x}
    + \frac{\gamma}{2} \left( \bar{n} + \frac{1}{2} \right)
    \boldsymbol{\nabla}
    \right) f (\boldsymbol{x},t),
\end{equation}
where $\boldsymbol{x} = (x, y)^{\rm T}$ and $\boldsymbol{\nabla}$ denotes the gradient with respect to $\boldsymbol{x}$.

This real-valued representation offers flexibility in interpretation, allowing for two distinct approaches depending on the parity of \(y\) under time-reversal symmetry. Considering the correspondence between $\alpha$ and the annihilation operator $\hat{a}$, it may be natural to interpret \(y\) as an odd-parity variable associated with momentum. Alternatively, \(y\) can be viewed as an additional position coordinate with even parity, leading the Fokker-Planck equation to describe a two-dimensional overdamped Brownian motion under linear forces.
In both interpretations, thermodynamic quantities corresponding to the heat flow and effective entropy production rate $\dot{\Pi}(t)$, which are defined in the original quantum system, can be identified for the classical effective dynamics described by Eq.~\eqref{eqA:real_FP_eq}.

In the case of the odd-parity interpretation, identifying a classical system governed by Eq.~\eqref{eqA:real_FP_eq} is unclear. Nonetheless, the formal definitions of thermodynamic quantities such as heat and entropy production, as established in stochastic thermodynamics, can still be applied.
The irreversible part of the probability current is given by
\begin{equation}
    \boldsymbol{J}^{\rm (o)}_{\rm irr} (\boldsymbol{x}, t)
    = - \frac{\gamma}{2} \left( \boldsymbol{x} + \left( \bar{n} + \frac{1}{2} \right) \boldsymbol{\nabla} \right) f(\boldsymbol{x},t)
\end{equation}
Using this, the entropy production rate is defined as
\begin{equation}
    \dot{\Sigma}^{\rm (o)} (t)
    \equiv
    \frac{2}{\gamma \left( \bar{n} + \frac{1}{2} \right)} \int \frac{\left|\boldsymbol{J}^{\rm (o)}_{\rm irr}(\boldsymbol{x},t) \right|^2}{f(\boldsymbol{x},t)} d^2\boldsymbol{x}.
\end{equation}
The irreversible current $\boldsymbol{J}^{\rm (o)}_{\rm irr} (\boldsymbol{x}, t)$ and the complex-valued phase-space current $J(\alpha,\alpha^*,t) = J_{\rm R} (\alpha, \alpha^*, t) + i J_{\rm I} (\alpha, \alpha^*, t)$, defined in Eq.~(11) of the main text, are linked by the relation $\boldsymbol{J}^{\rm (o)}_{\rm irr} (\boldsymbol{x}, t) =  - \sqrt{2}( J_{\rm R}(\alpha, \alpha^*, t), J_{\rm I}(\alpha, \alpha^*, t) )^{\rm T}$.
From this, we can see that the effective entropy production rate $\dot{\Pi}(t)$ in Eq.~(18) of the main text is identical to the entropy production rate $\dot{\Sigma}^{({\rm o})}(t)$ in this effective dynamics.

The definition of the Shannon entropy $\dot{S}(t)$ is independent of the parity of $y$.
Using the total entropy production and the Shannon entropy of the distribution, the heat flow can be identified as
\begin{equation}
    \dot{Q}^{\rm (o)}(t)
    = \tilde{T} \left( \dot{S} (t) - \dot{\Sigma}^{\rm (o)}(t) \right).
\end{equation}
After straightforward manipulations, it can be shown that the heat flow defined in the effective classical system is idential to that that in the original quantum dynamics:
\begin{equation}
    \dot{Q}^{\rm (o)}(t) = \int \boldsymbol{\Lambda}^{\rm T} (\boldsymbol{x}) \cdot \boldsymbol{J}^{\rm (o)}_{\rm irr}(\boldsymbol{x},t) d^2 \boldsymbol{x} = \dot{Q}(t),
\end{equation}
where $ \boldsymbol{\Lambda} (\boldsymbol{x}) = \hbar \omega \boldsymbol{x}$. This confirms that Eq.~(18) in the main text is equivalent to the second law of thermodynamics in this effective classical dynamics.
Furthermore, the entropic bound on the heat flow is expressed as~\cite{dechant2018entropic}
\begin{equation}
    \left( \dot{Q}^{\rm (o)} (t) \right)^2 \leq
    \frac{\gamma}{2} \left( \bar{n} + \frac{1}{2} \right) \langle |\boldsymbol{\Lambda}\bm{(}\boldsymbol{x}(t)\bm{)}|^2 \rangle \dot{\Sigma}^{\rm (o)} (t),
\end{equation}
whose time-integrated version corresponds to Eq.~(19) in the main text.

In the even-parity interpretation, the Fokker-Planck equation \eqref{eqA:real_FP_eq} describes as describing a two-dimensional overdamped Brownian particle subject to linear forces at a temperature $\tilde{T} = \hbar \omega (\bar{n} + 1/2)$.
To clarify this, we define effective coefficients $\tilde{k} = \hbar \omega$ (stiffness of a harmonic potential), $\tilde{\gamma} = 2 \hbar \omega / \gamma$ (damping coefficient) and $\tilde{\kappa} = 2 \hbar \omega^2 / \gamma$ (coefficient of a non-conservative force).
The Fokker-Planck equation can then be rewritten as
\begin{equation}
    \dot{f} (\boldsymbol{x},t)
    = - \boldsymbol{\nabla}^{\rm T} \cdot \left( \boldsymbol{\nu}^{\rm(e)} (\boldsymbol{x},t) f (\boldsymbol{x},t) \right),
\end{equation}
where $\bm{\nu}^{\rm(e)}$ is the mean local velocity defined as
\begin{equation}
    \boldsymbol{\nu}^{\rm (e)} (\boldsymbol{x},t)
    = \frac{1}{\tilde{\gamma} f(\boldsymbol{x},t)} \left( 
    \boldsymbol{F}^{\rm (e)}  (\boldsymbol{x})
    - \tilde{T}
    \boldsymbol{\nabla}
    \right) f(\boldsymbol{x},t),
\end{equation}
with a drift force
\begin{equation}
    \boldsymbol{F}^{\rm (e)} (\boldsymbol{x}) =
    \begin{pmatrix}
        - \tilde{k} & \tilde{\kappa} \\
        - \tilde{\kappa} & - \tilde{k}
    \end{pmatrix}
    \boldsymbol{x}.
\end{equation}
In this form, it becomes clear that the Fokker-Planck equation describes a two-dimensional overdamped Brownian motion under a harmonic potential $U(\bm{x}) = \tilde{k}\bm{x}^2/2$, accompanied by a non-conservative driving force $\bm{F}_{\rm nc}(\bm{x}) = \tilde{\kappa} (y, -x)^{\rm T}$.

Due to the non-conservative force, detailed balance is broken and the system relaxes to a nonequilibrium steady state, where entropy is continuously produced. In contrast, the original quantum system relaxes to an equilibrium state where the effective entropy production rate $\dot{\Pi}(t)$ vanishes. This indicates that the entropy production $\Sigma^{\rm (e)}(t)$ in this classical dynamics is not identical to $\Pi(t)$. Instead, the entropy production in this classical dynamics can be decomposed as
\begin{equation}
    \Sigma^{\rm (e)}(t)
    = \Sigma^{\rm (e)}_{\rm ex}(t) + \Sigma^{\rm (e)}_{\rm hk}(t),
\end{equation}
where the excess entropy production $\Sigma_{\rm ex}^{\rm (e)}(t)$ reflects the entropy production caused by the system's deviation from the stationary state, and the housekeeping entropy production $\Sigma_{\rm hk}^{\rm (e)}$ accounts for the contributions arising from the broken detailed balance~\cite{oono1998steady,esposito2010three,maes2014nonequilibrium,dechant2022geometric}. By decomposing the local mean velocity as $\nu^{\rm (e)}(t) = \nu^{\rm (e)}_{\rm ex}(t) + \nu^{\rm (e)}_{\rm hk}(t)$,
the rates of these entropy productions are given by $\dot{\Sigma}^{\rm (e)}_{\rm ex}(t) =  (\tilde{\gamma}/\tilde{T}) \langle |\boldsymbol{\nu}^{\rm (e)}_{\rm ex}(\boldsymbol{x},t)|^2 \rangle$ and $\dot{\Sigma}^{\rm (e)}_{\rm hk}(t) = (\tilde{\gamma}/\tilde{T}) \langle |\boldsymbol{\nu}^{\rm (e)}_{\rm hk}(\boldsymbol{x})|^2 \rangle$, where
\begin{equation}
    \boldsymbol{\nu}^{\rm (e)}_{\rm hk} (\boldsymbol{x})
    = \frac{1}{\tilde{\gamma} f^{\rm ss}(\boldsymbol{x})} \left( 
    \boldsymbol{F}^{\rm (e)}(\boldsymbol{x})
    - \tilde{T}
    \boldsymbol{\nabla}
    \right) f^{\rm ss} (\boldsymbol{x})
\end{equation}
with the steady-state distribution $f^{\rm ss} (\boldsymbol{x})$.
Using the expression for the steady-state distribution
\begin{equation}
    f^{\rm ss} (\boldsymbol{x})
    = \frac{\tilde{k}}{2 \pi \tilde{T}} e^{- \frac{\tilde{k}}{2 \tilde{T}} |\boldsymbol{x}|^2},
\end{equation}
we find that the housekeeping mean local velocity is
\begin{equation}
    \boldsymbol{\nu}_{\rm hk}^{\rm (e)} (\boldsymbol{x})
    = \frac{1}{\tilde{\gamma}}
    \left( \boldsymbol{F}^{\rm (e)}(\boldsymbol{x})
    + \tilde{k} \boldsymbol{x}
    \right),
\end{equation}
while the excess mean local velocity is
\begin{equation}
    \boldsymbol{\nu}^{\rm (e)}_{\rm ex} (\boldsymbol{x} , t)
    = - \frac{1}{\tilde{\gamma} f(\boldsymbol{x},t)} \left( \tilde{k} \boldsymbol{x} + \tilde{T} \nabla \right) f(\boldsymbol{x},t)=  - \sqrt{2} ( J_{\rm R}(\alpha, \alpha^*, t), J_{\rm I}(\alpha, \alpha^*, t) )^{\rm T}/f(\bm{x},t).
\end{equation}
This implies that the excess heat rate is given by~\cite{hatano2001steady,oono1998steady}
\begin{equation}
    \dot{Q}^{\rm (e)}_{\rm ex}(t) = \left \langle  \left( \boldsymbol{F}^{\rm (e)} \bm{(}\boldsymbol{x}(t)\bm{)} \right)^{\rm T} \cdot\boldsymbol{\nu}^{\rm (e)}_{\rm ex}\bm{(}\boldsymbol{x}(t),t\bm{)} \right \rangle = \left \langle \tilde{k} \boldsymbol{x}^{\rm T}(t) \cdot \boldsymbol{\nu}^{\rm (e)}_{\rm ex}\bm{(}\boldsymbol{x}(t),t\bm {)} \right \rangle = \dot{Q} (t),
\end{equation}
and the effective total entropy production rate $\dot{\Pi}(t)$ corresponds to the excess entropy production rate in this classical dynamics,
\begin{equation}
    \dot{\Pi}(t) = \dot{\Sigma}^{\rm (e)}_{\rm ex}(t) = \dot{S}(t) - \frac{\dot{Q}^{\rm (e)}_{\rm ex}(t)}{\tilde{T}} \geq 0.
\end{equation}
Thus, the heat flow and effective total entropy production rate in the original quantum system are directly related to the excess heat and excess entropy production in this classical dynamics.
Using the Cauchy-Schwarz inequality, we obtain
\begin{equation}
    \left( \dot{Q}^{\rm (e)}_{\rm ex}(t) \right)^2
    \leq \tilde{k}^2 \langle |\boldsymbol{x}(t)|^2 \rangle \langle |\boldsymbol{\nu}^{\rm (e)}_{\rm ex} \bm{(}\boldsymbol{x}(t),t\bm{)}|^2 \rangle,
\end{equation}
whose time-integrated version is equivalent to Eq.~(19) in the main text.
This relation is a variant of the entropic bound, where excess heat rate is bounded above by excess entropy production rate.

\section{power-efficiency trade-off for classical engines}

In this section, we derive a similar power-efficiency bound for a classical counter part of the quantum harmonic Otto engine. 
The Hamiltonian of the classical harmonic oscillator is given by $H(t) = p^2/(2m) + m\omega^2(t) x^2/2$.
The dynamics of adiabatic processes are governed by the Liouville equation associated with the Hamiltonian $H(t)$.
The relaxation dynamics during the isochoric processes can be modeled by a stochastic process governed by the Fokker-Planck equation
\begin{equation}\label{eq:classical_FP_eq_isochoric}
    \frac{\partial}{\partial t} f^{\rm cl}(x,p,t)
    = \mathcal{L}^{\rm cl}_\bullet f^{\rm cl}(x,p,t)
\end{equation}
with the differential operator
\begin{equation}
    \mathcal{L}^{\rm cl}_\bullet 
    = - \frac{p}{m} \frac{\partial}{\partial x} 
    + \frac{\partial}{\partial p} \left( m \omega_\bullet^2 x + \frac{\gamma^{\rm cl}}{m} p + \gamma^{\rm cl} T_\bullet \frac{\partial}{\partial p}  \right).
\end{equation}
The damping coefficient $\gamma^{\rm cl}$ of the thermal reservoir is a free parameter which is entirely independent of $\gamma$ from the quantum Lindblad operator.
It is worth noting that these dynamics are fundamentally different from the effective classical dynamics in the previous section, which is derived by transforming variables from the complex-valued Fokker-Planck equation.

The rate of average heat dissipation is given by
\begin{equation}
\begin{aligned}
    \dot{Q}^{\rm cl}_\bullet
    & = \int dx \int dp ~ J_{\rm irr}^{\rm cl}(x,p,t) \frac{p}{m} \\
    & = \frac{\gamma^{\rm cl}}{m} \left(
    T_\bullet - \int dx \int dp ~ \frac{p^2}{m} f^{\rm cl}(x,p,t)
    \right)
\end{aligned}
\end{equation}
where the irreversible part of probability current is given by~\cite{gardiner2009stochastic}
\begin{equation}
    J^{\rm cl}_{\rm irr}(x,p,t) = -\gamma^{\rm cl} \left( \frac{p}{m} + T_\bullet \frac{\partial}{\partial p} \right) f^{\rm cl}(x,p,t).
\end{equation}
Owing to the analogy of this expression with Eq.~(15) in the main text, and applying the Cauchy-Schwartz inequality, we obtain
\begin{align}
    \left( Q^{\rm cl}_{\rm h} \right)^2
     &\leq \left( \frac{1}{m^2} \int_{\mathcal{I}_{\rm h}} dt \int dx \int dp ~p^2 f^{\rm cl}(x,p,t) \right) \left( \int_{\mathcal{I}_{\rm h}} dt \int dx \int dp \frac{\bm{(}J^{\rm cl}_{\rm irr} (x,p,t)\bm{)}^2}{f^{\rm cl} (x,p,t)} \right) \\
     &= T_{\rm h} \Sigma_{\rm h} \left( \frac{\gamma^{\rm cl}}{m} T_{\rm h}  \Delta t(\mathcal{I_{\rm h}}) - Q_{\rm h} \right).
\end{align}
where $\Sigma_{\rm h}$ denotes the total entropy produced during the process $\mathcal{I}_{\rm h}$.
This inequality has the same form as Eq.~(19) in the main text. Following the same steps as in the main text, we obtain
\begin{equation}
    P \leq \frac{\gamma^{\rm cl} T_{\rm h} {\rm min}  \{ \Delta t(\mathcal{I}_{\rm h}), \Delta t(\mathcal{I}_{\rm c})\} }{m \tau}
    \frac{\eta(\eta_C - \eta)}{1-\eta}.
\end{equation}
Thus, the power-efficiency trade-off for the classical harmonic Otto heat engine takes the same form as that for the quantum engine, but with the parameters $\{ \gamma, \tilde{T}_{\rm h}, \tilde{\eta} \}$ replaced by $\{ \gamma^{\rm cl}/m, T_{\rm h}, \eta_C \}$.

\section{Generalized quasi-probability distributions}

In this section, we explain why we focused exclusively on the Wigner function in our derivation by demonstrating that the approach is incompatible with other quasi-probability representations.
For completeness, we first briefly review the definition of  generalized quasi-probability distributions.
These distributions share a common property: the correspondence between the moments of quantum operators and their associated classical fields.
For instance, the Wigner function satisfies
\begin{equation}
    \int (\alpha^*)^m \alpha^n W(\alpha, \alpha^*) d^2 \alpha
    =
    {\rm tr}{\bm (} \hat{\rho} \{ ( \hat{a}^\dagger)^m \hat{a}^n \}_{\rm sym}  {\bm )},
\end{equation}
where $\{ ( \hat{a}^\dagger)^m \hat{a}^n \}_{\rm sym} = (-i \partial_\xi)^n (-i \partial_{\xi^*})^m e^{-i (\xi \hat{a} + \xi^* \hat{a}^\dagger)} |_{(\xi, \xi^*)=(0,0)}$ represents the symmetrically ordered product~\cite{puri2001mathematical}.
This relation shows that other quasi-probability functions can be constructed by establishing correspondence to the quantum joint moments using alternative ordering conventions.
Two commonly used orderings are normal ordering and anti-normal ordering, expressed as
$\{ ( \hat{a}^\dagger)^m \hat{a}^n \}_{\rm norm} = (\hat{a}^\dagger)^m \hat{a}^n$
and
$\{ ( \hat{a}^\dagger)^m \hat{a}^n \}_{\rm anti-norm} = \hat{a}^n (\hat{a}^\dagger)^m $,
respectively.
For these orderings, the corresponding quasi-probability distributions can be defined by taking the Fourier transform of their characteristic functions. However, these distributions are often expressed in a more compact form using the coherent state $|\alpha \rangle$, which is the eigenstates of the annihilation operator satisfying $\hat{a} |\alpha\rangle = \alpha |\alpha \rangle$.
In the case of normal ordering, the Glauber P-function $P(\alpha,\alpha^*)$ is given by the relation $\hat{\rho} = \int P(\alpha,\alpha^*) |\alpha \rangle \langle \alpha | d^2\alpha$.
For anti-normal ordering, the Husimi Q-function $Q(\alpha,\alpha^*)$ can be expressed as $Q(\alpha,\alpha^*) = \langle \alpha | \hat{\rho} | \alpha \rangle/\pi$~\cite{gardiner2004quantum}.

In general, one may consider a generalized ordering of exponentiated operators to define the characteristic function. With any number of arbitrary complex coefficients $\beta_i$ and $\gamma_i$ for $i = 1, ..., N$, the most general form of product can be expressed as
\begin{equation}
    \{ e^{-i(\xi \hat{a} + \xi^* \hat{a}^\dagger)} \}_{\rm gen} = e^{-i\beta_1 \xi \hat{a}} e^{-i\gamma_1 \xi^* \hat{a}^\dagger} e^{-i\beta_2 \xi \hat{a}} e^{-i\gamma_2 \xi^* \hat{a}^\dagger} \cdots e^{-i\beta_N \xi \hat{a}} e^{-i\gamma_N \xi^* \hat{a}^\dagger}.
\end{equation}
By repeatedly applying the Baker-Hausdorff formula, this expression can be rewritten compactly as
\begin{equation}
    \{ e^{-i(\xi \hat{a} + \xi^* \hat{a}^\dagger)} \}_{\rm gen} 
    = e^{-\frac{s}{2}|\xi|^2} e^{-i(\xi\hat{a}+\xi^*\hat{a}^\dagger)}
    \equiv
    \{ e^{-i(\xi \hat{a} + \xi^* \hat{a}^\dagger)} \}_s
\end{equation}
where the complex coefficient $s$ is given by $s = 1 - 2\sum_{j=2}^N\sum_{k=1}^{j-1} \beta_j \gamma_k$.
This result shows that any ordering of the operators can be characterized by a single complex parameter $s$.
By taking the Fourier transform of the corresponding characteristic function, we can define a generalized quasi-probability distribution as~\cite{cahill1969ordered,cahill1969density,puri2001mathematical}.
\begin{equation}\label{eq:def_quasi_prob}
\begin{aligned}
    f_s(\alpha,\alpha^*)
    & = \frac{1}{\pi^2}\int {\rm tr} \left( \hat{\rho} \{ e^{-i(\xi\hat{a}+\xi^*\hat{a}^\dagger)} \}_s \right) e^{i(\xi\alpha + \xi^*\alpha^*)} d^2\xi \\
    & = \frac{1}{\pi^2} \int e^{-\frac{s}{2}|\xi|^2} {\rm tr} \left( \hat{\rho} e^{-i(\xi\hat{a}+\xi^*\hat{a}^\dagger)} \right) e^{i(\xi\alpha + \xi^*\alpha^*)} d^2\xi.
\end{aligned}
\end{equation}
The identities $e^{-i(\xi\hat{a}+\xi^*\hat{a}^\dagger)} = e^{|\xi|^2/2}e^{-i\xi\hat{a}}e^{-i\xi^*\hat{a}^\dagger} = e^{-|\xi|^2/2}e^{-i\xi^*\hat{a}^\dagger}e^{-i\xi\hat{a}}$ imply $s= -1, 0$, and $1$ correspond to the Glauber $P$-function, the Wigner function, and the Husimi $Q$-function, respectively~\cite{puri2001mathematical,gardiner2004quantum}.
Once a quasi-probability distribution for a given $s$ exists, the definition \eqref{eq:def_quasi_prob} ensures that it is normalized by $\int f_s(\alpha,\alpha^*) d^2\alpha = 1$.

During the isochoric processes, the time evolution of the generalized quasi-probability distribution is governed by the same complex-valued Fokker-Planck equation as in the main text but with a different probability current
\begin{equation}
    J_{\bullet,s}(\alpha,\alpha^*,t)
    = \frac{\gamma \alpha}{2} f_s(\alpha,\alpha^*,t) \\
    + \frac{\gamma}{2} \left( \bar{n}_\bullet + \frac{s+1}{2} \right)
    \frac{\partial f_s(\alpha,\alpha^*,t)}{\partial \alpha^*}.
\end{equation}
It implies that different values of $s$ result in different effective temperatures for the effective classical dynamics. For a physically meaningful mapping, we focus exclusively on cases where $s$ is a real value.
During the adiabatic processes, the operators $\hat{a}(t)$ and $\hat{a}^\dagger(t)$ explicitly depend on time. Therefore, we must redefine the generalized quasi-probability distribution as
\begin{equation}
    f_s(\alpha,\alpha^*,t)
    = \frac{1}{\pi^2} \int e^{-\frac{s}{2}|\xi|^2} {\rm tr} \{ \hat{\rho}(t) e^{-i(\xi\hat{a} (t) +\xi^*\hat{a}^\dagger (t))} \} e^{i(\xi\alpha + \xi^*\alpha^*)} d^2\xi.
\end{equation}
From a relation
\begin{equation}
    \partial_t \hat{a} (t) = \frac{\dot{\omega}(t)}{2 \omega(t)}
    \hat{a}^\dagger (t)
\end{equation}
and the Sneddon's formula
\begin{equation}
    \partial_t e^{\hat{A}(t)}
    = \int_0^1 du ~ e^{u \hat{A}(t)} (\partial_t \hat{A}) e^{(1-u) \hat{A}(t)},
\end{equation}
we obtain
\begin{equation}
    \partial_t e^{-i (\xi \hat{a}(t) + \xi^* \hat{a}^\dagger (t)}
    = \frac{\dot{\omega}(t)}{2 i \omega(t)}
   e^{-i (\xi \hat{a}(t) + \xi^* \hat{a}^\dagger (t))}
   \left( \xi \left( \hat{a}^\dagger (t) + i \frac{\xi}{2}\right)
   + \xi^* \left( \hat{a} + i \frac{\xi^*}{2} \right)\right).
\end{equation}
This yields the equation for $f_s (\alpha, \alpha^*,t)$ during the adiabatic process as
\begin{equation}\label{eq:L_As}
    \frac{\partial}{\partial t} f_s(\alpha,\alpha^*,t)
    = -i\omega(t) \left(-\frac{\partial}{\partial \alpha} \alpha
    + \frac{\partial}{\partial \alpha^*} \alpha^* \right)
    f_s(\alpha,\alpha^*,t)
    + \frac{\partial J_{\mathcal{A},s}^*(\alpha, \alpha^*, t)}
    {\partial \alpha}
    + \frac{\partial J_{\mathcal{A},s}(\alpha, \alpha^*, t)}
    {\partial \alpha^*}
\end{equation}
where
\begin{equation}
    J_{\mathcal{A},s}(\alpha,\alpha^*,t)
    = - \frac{\dot{\omega}(t)}{2 \omega(t)} \left(
    \alpha + \frac{s}{2} \frac{\partial}{\partial \alpha^*}
    \right)
    f_s(\alpha,\alpha^*,t) .
\end{equation}
This relation shows that for $s\neq0$, the time evolution operator includes a second-order derivative term, leading to changes in the Shannon entropy of the quasi-probability distribution.
As a result, after completing a cycle, the Shannon entropy change does not vanish and the effective total entropy production can not be simply expressed in Clausius form. 
Therefore, we exclusively adhere to the choice of $s=0$ in the main text.

\section{Equality conditions}
The equality of the bound can be achieved by satisfying two conditions. 
First, to satisfy the equality of the Cauchy-Schwartz inequality applied in Eq.~(19) of the main text, the local mean velocity $J(\alpha,\alpha^*,t)/f(\alpha,\alpha^*,t)$ must be proportional to the phase variable $\alpha$ throughout the isochoric process.
Using the definition of the current in Eq.~(11) of the main text, this condition can be shown to be equivalent to $\partial f(\alpha,\alpha^*,t)/\partial \alpha^* = \mu \alpha f(\alpha,\alpha^*,t)$ where $\mu$ is an arbitrary constant.
When the heat engine reaches the periodic steady-state, the Wigner function is given by a Gaussian distribution,
\begin{equation}\label{eq:Gaussian_Wigner}
    f(\alpha,\alpha_*,t)
    = \frac{1}{\pi \sqrt{N^2(t) - |M|^2(t)}} \exp\left( - \frac{2N(t)|\alpha|^2 - M(t) \alpha^{*2} - M^*(t) \alpha^2}{2 \bm{(} N^2(t) - |M|^2(t) \bm{)}} \right),
\end{equation}
where $N(t) = \int |\alpha|^2 f(\alpha,\alpha_*,t) d^2 \alpha$ and $M(t) = \int \alpha^2 f(\alpha,\alpha_*,t) d^2 \alpha$.
Substituting this expression into the equality condition gives
$-\bm{(} N(t) \alpha - M(t) \alpha^* \bm{)}/\bm{(}N^2(t) - |M|^2(t) \bm{)} = \mu \alpha$. Thus, the equality holds when $|M(t)|$ is significantly smaller than $N(t)$ throughout the isochoric process.
Physically, this condition implies that minimizing coherence is necessary to achieve high power.
This can be understood by noting that $M$ serves as a measure of coherence generated in the working substance.
From the definition of the Wigner function, $M$ can be expressed as ${\rm tr} (\hat{\rho} \hat{a}^2 )$.
This is consistent with the common understanding that coherence is detrimental to the performance of a heat engine, as it acts a form of internal friction~\cite{kosloff2002discrete,thomas2014friction,alecce2015quantum,ccakmak2017irreversible}.

The second condition arises from the inequality $\Pi_{\rm h} \leq \Pi_{\rm h} + \Pi_{\rm c}$ for $\Delta t(\mathcal{I}_{\rm h}) < \Delta t(\mathcal{I}_{\rm c})$, or $\Pi_{\rm c} \leq \Pi_{\rm h} + \Pi_{\rm c}$ for $\Delta t(\mathcal{I}_{\rm h}) > \Delta t(\mathcal{I}_{\rm c})$.
If the condition for minimizing coherence is satisfied and one of time durations of the isochoric processes is much smaller than others, for example $\Delta t(\mathcal{I}_{\rm h}) \ll \Delta t(\mathcal{I}_{\rm c})$, one can expect that the periodic steady-state of the system will closely resemble the equilibrium state at temperature $T_{\rm c}$.
In this case, the entropy production during the longer isochoric process becomes negligible $\Pi_{\rm c} \approx 0$, resulting in $\Pi_{\rm h} \approx \Pi_{\rm h} + \Pi_{\rm c}$. The same reasoning applies to the opposite scenario, where $\Delta t(\mathcal{I}_{\rm h}) \gg \Delta t(\mathcal{I}_{\rm c})$.
In conclusion, the bound is attained when $\Delta t(\mathcal{A})$ is much larger than the characteristic timescale of the protocol, and either $\Delta t(\mathcal{I}_{\rm h}) \gg \Delta t(\mathcal{I}_{\rm c})$ or $\Delta t(\mathcal{I}_{\rm h}) \ll \Delta t(\mathcal{I}_{\rm c})$.

\section{Engine with quasistatic adiabatic processes}

The Gaussian form of the Wigner distribution in Eq.~\eqref{eq:Gaussian_Wigner} implies that the periodic steady state of the system is fully characterized by the time evolution of two moments $N(t)$ and $M(t)$.
To derive the coupled closed equations for $N(t)$ and $M(t)$, it is simpler to describe the dynamics using the corresponding Langevin equations.
The Langevin equation for isochoric processes is given by
\begin{equation}
    \dot{\alpha}(t) = - \left( \frac{\gamma}{2} + i\omega_\bullet \right) \alpha(t) 
    + \sqrt{\gamma \left( \bar{n}_\bullet + \frac{1}{2} \right)} \xi(t),
\end{equation}
which is analytically solvable since it is linear in $\alpha(t)$ with time-independent coefficients.
The solution of the Langevin equation is
\begin{equation}
\begin{aligned}
    \alpha(t) 
    & = e^{-(\gamma/2 + i \omega_\bullet)(t-t_0)}\alpha(t_0) \\
    & ~~~ + \sqrt{\gamma \left( \bar{n}_\bullet + \frac{1}{2} \right)}\int_{t_0}^t d\tau ~ e^{-(\gamma/2 + i\omega_\bullet)(t-\tau)} \xi(\tau).
\end{aligned}
\end{equation}
This connects the second moments at the beginning and end of each isochoric process as (see Fig.~1 of the main text)
\begin{equation}\label{eq:NM_eq_ischoric}
\begin{aligned}
    N_2 & = e^{-\gamma \Delta t(\mathcal{I}_h)} N_1
    + \left( \bar{n}_h + \frac{1}{2} \right) \left( 1 - e^{-\gamma \Delta t(\mathcal{I}_h)} \right), \\
    M_2 & = e^{-(\gamma + 2i \omega_h) \Delta t(\mathcal{I}_h)} M_1, \\
    N_4 & = e^{-\gamma \Delta t(\mathcal{I}_c)} N_3
    + \left( \bar{n}_c + \frac{1}{2} \right) \left( 1 - e^{-\gamma \Delta t(\mathcal{I}_c)} \right), \\
    M_4 & = e^{-(\gamma + 2i \omega_c) \Delta t(\mathcal{I}_c)} M_3.
\end{aligned}
\end{equation}

The deterministic equation for adiabatic processes is
\begin{equation}\label{eq:deterministic_eq_adiabatic}
    \dot{\alpha}(t) = -i\omega(t) \alpha(t) + \frac{\dot{\omega}(t)}{2\omega(t)} \alpha^*(t),
\end{equation}
Finding a closed-form solution for this equation analytically is generally challenging, except under special protocols of $\omega(t)$.
To proceed, We consider a quasistatically slow adiabatic process during which $\dot{\omega}(t)/\omega(t) \ll \omega(t)$.
Neglecting the second term on the right-hand side of Eq.~\eqref{eq:deterministic_eq_adiabatic}, we obtain
\begin{equation}
    \alpha(t) = e^{-i \int_{t_0}^t \omega(\tau) d\tau} \alpha(t_0),
\end{equation}
which connects the second moments at the beginning and end of each adiabatic process as (see Fig.~1 of the main text)
\begin{equation}\label{eq:NM_eq_adiabatic_quasistatic}
\begin{aligned}
    N_3 = N_2, ~~ M_3 = e^{-2i\int_{\mathcal{A}_{h\to c}} \omega(t) dt} M_2,  \\
    N_1 = N_4, ~~ M_1 = e^{-2i\int_{\mathcal{A}_{c\to h}} \omega(t) dt} M_4.
\end{aligned}
\end{equation}
This result is consistent with the quantum adiabatic theorem, which states that the population of occupied energy levels does not change during a quasistatic adiabatic process.
By solving the coupled equations in Eqs.~\eqref{eq:NM_eq_ischoric} and \eqref{eq:NM_eq_adiabatic_quasistatic} to find the cyclic steady-state values of the second moments, we obtain simple expressions for the efficiency and power of the slow engine as follows~\cite{rezek2006irreversible}:
\begin{equation}\label{eq:efficiency_quasistatic}
    \eta_q = 1 - \frac{\omega_c}{\omega_h}
\end{equation}
and
\begin{equation}\label{eq:power_quasistatic_appendix}
\begin{aligned}
    P_q = \frac{\hbar(\omega_h-\omega_c)(\bar{n}_h - \bar{n}_c)}{\tau} \frac{ ( 1 - e^{-\gamma \Delta t (\mathcal{I}_h)} ) ( 1 - e^{-\gamma \Delta t (\mathcal{I}_c)} )}{1 - e^{-\gamma \bm{(} \Delta t(\mathcal{I}_h) + \Delta t(\mathcal{I}_c) \bm{)} })}.
\end{aligned}
\end{equation}
Combining Eqs.~\eqref{eq:efficiency_quasistatic} and \eqref{eq:power_quasistatic_appendix} leads to Eq.~(7) of the main text.

\section{Details of Numerical simulation}

\begin{figure}[t]
\centering
\includegraphics[width=\columnwidth]{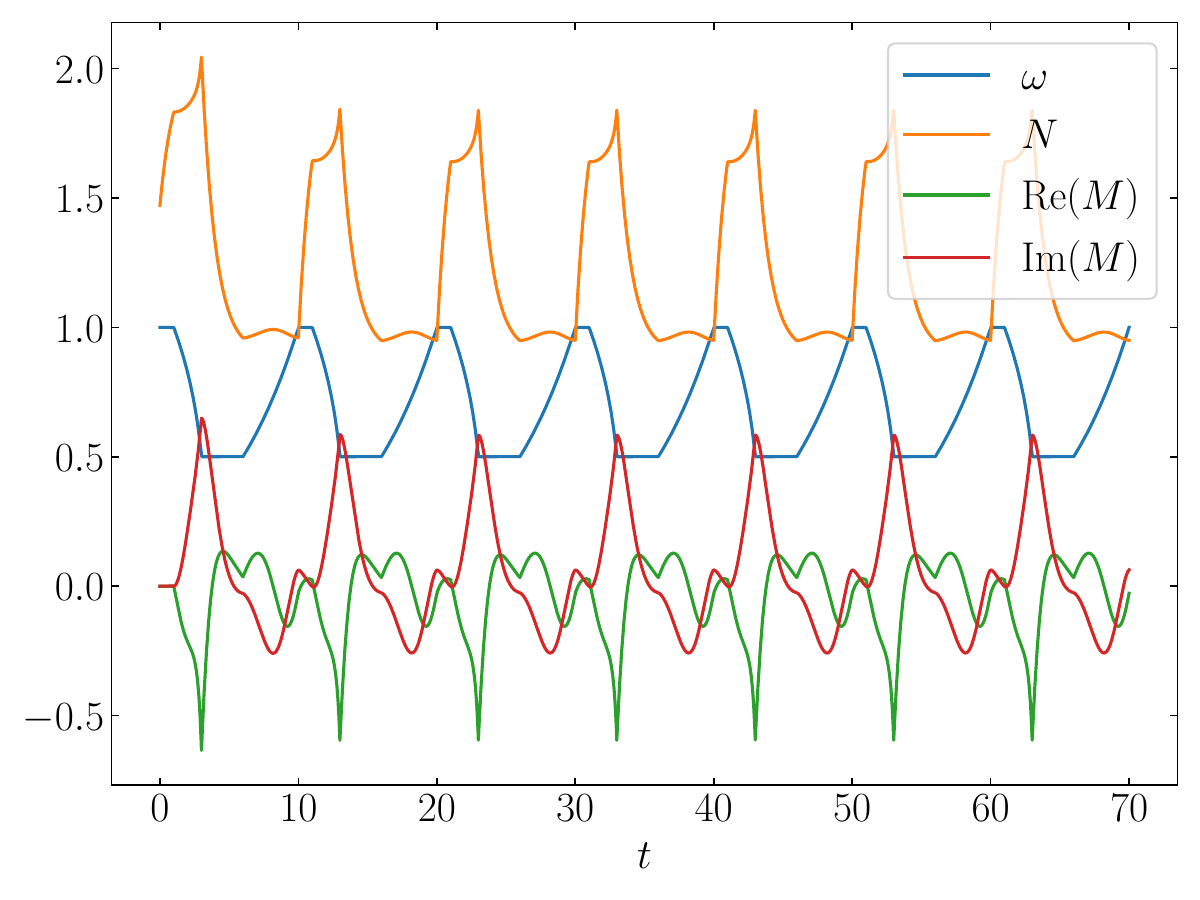}
\caption{An example of a combined protocol and the resulting time evolution of $N(t)$ and $M(t)$ are depicted.
The frequency changes during the adiabatic processes satisfy $\dot{\omega}(t) \propto \omega^{-2}(t)$ for $\mathcal{A}_{{\rm h}\to {\rm c}}$ and $\dot{\omega}(t) \propto \omega(t)$ for $\mathcal{A}_{{\rm c}\to {\rm h}}$.
The initial values of $N(t)$ and $M(t)$ are set as $N(0) = (\bar{n}_h + \bar{n}_c+1)/2$ and $M(0) = 0$.
The durations of the processes are chosen as $\Delta t(\mathcal{I}_h) = 1$, $\Delta t(\mathcal{A}_{h\to c}) = 2$, $\Delta t(\mathcal{I}_c) = 3$, and $\Delta t(\mathcal{A}_{c\to h}) = 4$, with system parameters $(\hbar \omega_h, \hbar \omega_c, \beta_h, \beta_c, \gamma) = (1, 0.5, 0.5, 2.5, 1)$.}
\label{fig:combined_protocol}
\end{figure}

In this section, we first explain the details of numerical simulations and then display the results for different types of protocols.
As mentioned in the main text, the heat and work are determined by the second-order moments, $N(t) = \int |\alpha|^2 f(\alpha,\alpha^*,t) d^2\alpha$, $M(t) = \int \alpha^2 f(\alpha,\alpha^*,t) d^2\alpha$, and $M^*(t)$ .
During adiabatic processes, the time-evolution equations of the second moments are coupled as
\begin{equation}\label{eq:second_moments_adiabatic}
    \frac{d}{dt} \begin{pmatrix}
    N(t) \\ M(t) \\ M^*(t)
    \end{pmatrix}
    = 
    \mathsf{\Lambda}(t) \cdot \begin{pmatrix}
    N(t) \\ M(t) \\ M^*(t)
    \end{pmatrix}
\end{equation}
with the matrix
\begin{equation}
    \mathsf{\Lambda}(t) = \frac{1}{\omega(t)}\begin{pmatrix}
        0 & \frac{\dot{\omega}(t)}{2} & \frac{\dot{\omega}(t)}{2}\\
        \dot{\omega}(t) & -2i\omega^2(t) & 0 \\
        \dot{\omega}(t) & 0 & 2i\omega^2(t)
    \end{pmatrix}.
\end{equation}
In contrast, the moments remain decoupled during isochoric processes, and their time evolutions are governed by
\begin{equation}\label{eq:second_moments_isochoric}
\begin{aligned}
    \frac{dN(t)}{dt}
    & = \gamma \left( \bar{n}_\bullet + \frac{1}{2} - N(t) \right), \\
    \frac{d M(t)}{dt} 
    & = -(\gamma + 2i\omega_\bullet) M(t).
\end{aligned}
\end{equation}

Assuming the heat engine operates in the cyclic steady state, we denote the values of ${\bm (} N(t),M(t) {\bm )}$ at the beginning of each process by $(N_1,M_1), (N_2, M_2), (N_3,M_3)$, and $(N_4,M_4)$ (see Fig.~1 of the main text).
Once $N_i$ is computed, the amounts of heat, $Q_{\rm h}$ and $Q_{\rm c}$, can be obtained as $Q_{\rm h} = \hbar \omega_{\rm h} ( N_2 - N_1)$ and $Q_{\rm c} = \hbar \omega_{\rm c} ( N_4 - N_3)$. By solving Eq.~\eqref{eq:second_moments_isochoric}, the heat can also be expressed as
\begin{equation}
\begin{aligned}
    Q_h & = \hbar\omega_h \left( \bar{n}_h + \frac{1}{2} - N_1 \right) \left( 1 - e^{-\gamma \Delta t(\mathcal{I}_h)} \right), \\
    Q_c & = -\hbar\omega_c \left( \bar{n}_c + \frac{1}{2} - N_3 \right) \left( 1- e^{-\gamma \Delta t(\mathcal{I}_c)} \right).
\end{aligned}
\end{equation}
The total amount of extracted work can then be obtained from the condition of energy conservation, $W_{\rm ext} = Q_h - Q_c$.
In contrast to the isochoric processes, solving the coupled equations for the adiabatic processes in \eqref{eq:second_moments_adiabatic} analytically for a general protocol $\omega(t)$ is challenging, necessitating the use of numerical methods to calculate the power and efficiency.
Nevertheless, when $\dot{\omega}(t)$ is proportional to $\omega^2(t)$, the time-dependence can be factored out of the matrix $\mathsf{\Lambda}(t)$, making an analytical solution feasible, albeit complex~\cite{kosloff2017quantum,lee2020finite,lee2021quantumness}.
By numerically calculating the analytical solution, we evaluated $N_i$.
To obtain the steady-state values, we first repeated the cycle seven times, starting with the initial values $N_1 = (n_{\rm h} + n_{\rm c})/2$ and $M=0$. 
We then calculated the standard deviation of $N_i$ over the last five cycles, as the system typically reaches the steady-state after two cycles as shown in Fig.~\ref{fig:combined_protocol}.
Finally,
to produce Figs. 2 and 3 in the main text,
we averaged $N_i$ over the last five cycles only when the standard deviation was below a predefined small threshold.

For further demonstrations, we considered other protocols defined as
\begin{equation}
    \omega_m(t) = \left( \omega_i^{1-m} + \left( \omega_f^{1-m} - \omega_i^{1-m} \right) \frac{t}{\Delta t(\mathcal{A})} \right)^{\frac{1}{1-m}},
\end{equation}
where the initial or final value of the natural frequencies, $\omega_i$ or $\omega_f$, is set to $\omega_1$ or $\omega_2$, depending on the specific adiabatic process. This protocol satisfies the relation $\dot{\omega}(t) \propto \omega^m(t)$. For $m=1$, the protocol becomes
\begin{equation}
    \omega_1(t) = \lim_{m \rightarrow 1} \omega_m(t)
    = \omega_i \left( \frac{\omega_f}{\omega_i} \right)^{\frac{t}{\Delta t(\mathcal{A})} }.
\end{equation}
To speed up the simulation, we use the analytical solution in Eq.~\eqref{eq:NM_eq_ischoric} for the isochoric processes to calculate the second-order moments at the end of the processes. For the adiabatic processes, we represent $M(t)$ in terms of its real and imaginary parts, $M(t) = M_R(t) + M_I (t) i$, and solve the coupled equations
\begin{equation}
    \frac{d}{dt} \begin{pmatrix}
    N(t) \\ M_R(t) \\ M_I(t)
    \end{pmatrix}
    =
    \frac{1}{\omega(t)}
    \begin{pmatrix}
        0 & \frac{\dot{\omega}(t)}{2} & 0 \\
        \dot{\omega}(t) & 0 & 2 \omega^2(t) \\
        0 & - 2 \omega^2 (t) & 0
    \end{pmatrix}
    \cdot \begin{pmatrix}
    N(t) \\ M_R(t) \\ M_I(t)
    \end{pmatrix}.
\end{equation}
For a given choice of protocols, we solve this equation numerically by using the forth-order Runge-Kutta method.
\begin{figure}[t]
\centering
\includegraphics[width=\columnwidth]{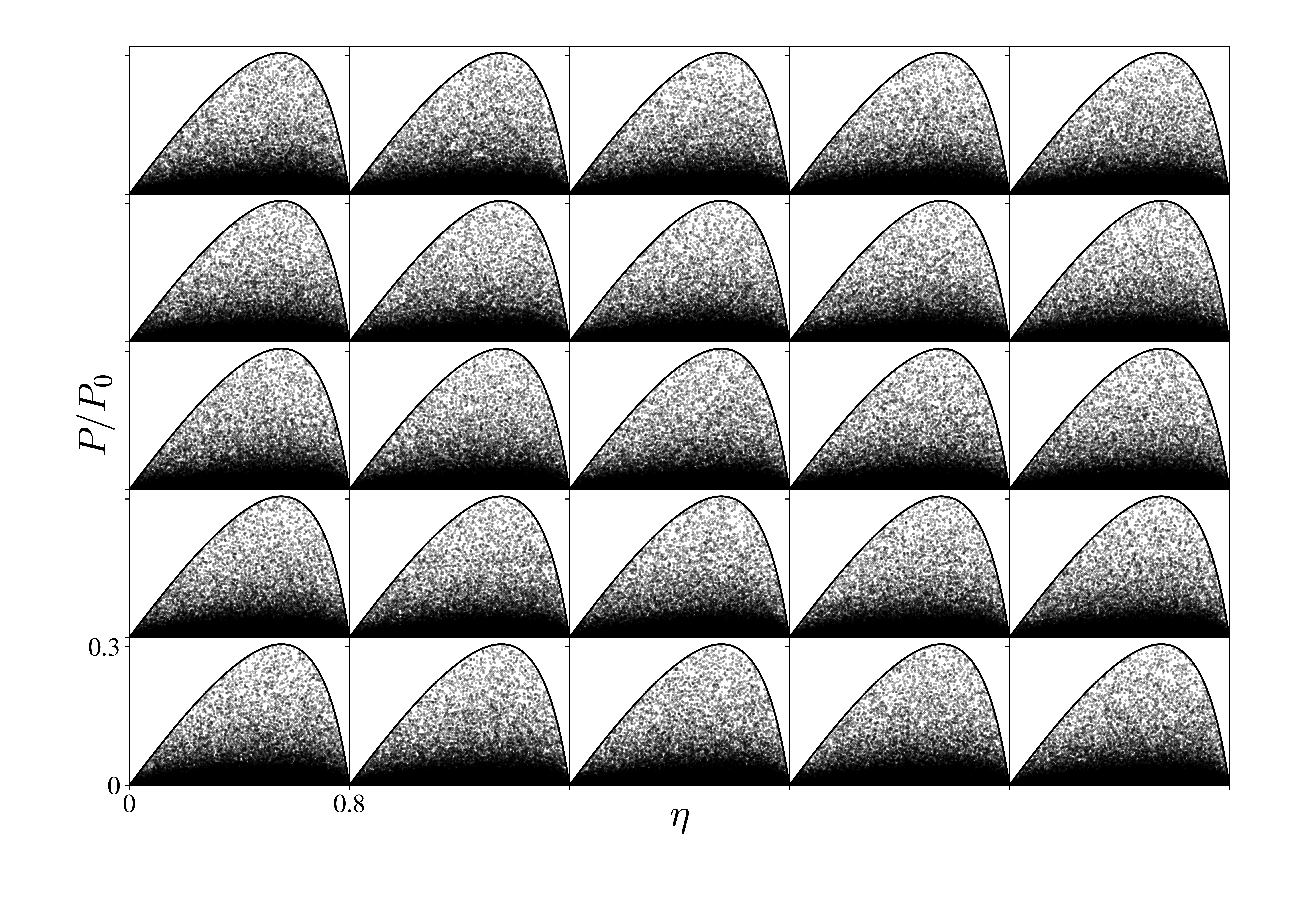}
\caption{The scatter plot shows the power-efficiency trade-off relation for various protocols. Each point represents results obtained for a set of randomly selected system parameters. Each column and row correspond to different protocols for the adiabatic processes. Five different protocols are used, characterized by the exponent $m = -2, -1, 0, 1,$ and $2$ for the adiabatic compression (left to right) and adiabatic expansion (top to bottom). Each plot depicts $10^5$ points.}
\label{fig:trade_offs}
\end{figure}

Figure~\ref{fig:trade_offs} shows the results obtained from the $5 \times 5$ combinations of protocols for each adiabatic process. For each adiabatic process, the protocol was assigned as $\omega_m (t)$ with $m=-2, -1, 0, 1,$ and $2$. For all cases, we fixed $\tilde{\eta} = 0.8$ and $\gamma = 1$. The parameters $\eta_{\rm O}$, $\omega_{\rm h}$, and $\bar{n}_{\rm h}$ were randomly sampled from the ranges $[0, 0.8]$, $[0, 500]$, $[0, 10]$, respectively. These values were then used to determine $\omega_c$ $T_{\rm h}$, and $T_{\rm c}$. The time duration of the adiabatic processes was fixed at $\Delta t(\mathcal{A}_i)=100$, while the durations of the isochoroic processes were randomly selected from $[0, 100]$. 

\begin{figure}[t]
\centering
\includegraphics[width=\columnwidth]{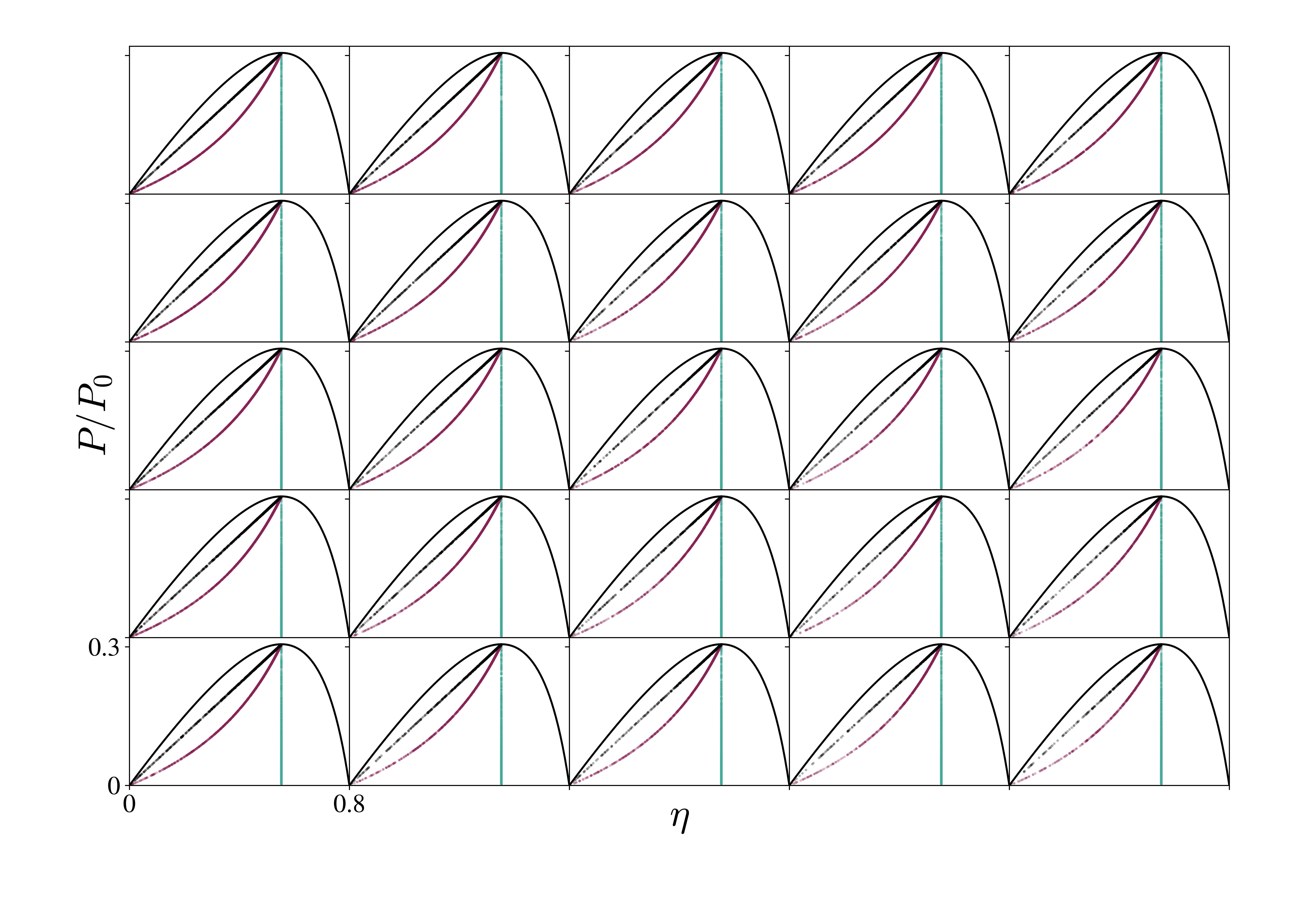}
\caption{The scatter plot shows the validity of the equality conditions. Each column and row correspond to the same protocols as in Fig. 2. Cyan symbols indicate the data obtained for the quasi-static adiabatic processes, while red and blue symbols present data for the second condition where $\Delta t(\mathcal{I}_{\rm h}) 
\ll \Delta t(\mathcal{I}_{\rm c})$ and $\Delta t(\mathcal{I}_{\rm c}) 
\ll \Delta t(\mathcal{I}_{\rm h})$ respectively. All plots consistently depict $10^4$ points for each type of symbols.}
\label{fig:EQ_conditions}
\end{figure}

Figure~\ref{fig:EQ_conditions} shows the validation of the equality conditions. The grid of plots displays data obtained from different combinations of protocols. Here, we set $\tilde{\eta} = 0.8$ $\gamma = 1$, and $\eta_{\rm O} = 1 - \sqrt{1 - \tilde{\eta}}$. The values of $\omega_{\rm h}$ and $\bar{n}_{\rm h}$ were randomly selected from the ranges $[0.1, 10]$ and $[0, 10]$, respectively. Cyan points represent data obtained for the quasi-static adiabatic processes. To meet this condition, we set $\Delta t(\mathcal{A}_i) = 100000$, while $\Delta t(\mathcal{I}_i)$ was sampled from $[0, 100]$. Red and blue points represent data obtained under the second condition of equality, where $\Delta t(\mathcal{A}_i)$ was sampled from $[0, 100]$, and $\Delta t(\mathcal{I}_i)$ was sampled from $[0, 100]$ or $[0, 0.01]$, respectively.

\bibliographystyle{apsrev}
\bibliography{main}